%% file: aa57425-25.tex
\documentclass{aa}

\usepackage{graphicx}
\usepackage{txfonts}
\usepackage{lastpage}
\usepackage{afterpage}
\usepackage{booktabs}
\usepackage{tabularx}
\usepackage{orcidlink}
\usepackage{pdflscape} 
\usepackage{hyperref}
\usepackage{url}

\usepackage{gensymb}
\usepackage{sidecap} 

\usepackage[dvipsnames]{xcolor}
\usepackage{booktabs}

\usepackage{subcaption}
\usepackage[skip=0pt]{caption}
\begin{document}

   \title{Coordinated Space- and Ground-based Monitoring of Accretion Bursts in a Protoplanetary Disk:\\ The Orbital and Accretion Properties of DQ Tau\thanks{Based on observations collected at the European Organization for Astronomical Research in the Southern Hemisphere under ESO programs 114.2799.001, 114.27MG.001, and 082.C-0218(A).}}

 \author{
    Hala Alqubelat\orcidlink{0009-0002-4535-1704}\inst{1},
    Carlo F. Manara\orcidlink{0000-0003-3562-262X}\inst{1},
    Justyn Campbell-White\orcidlink{0000-0002-3913-3746}\inst{1},
    Monika G. Petr-Gotzens\inst{1},\\
    Benjamin M. Tofflemire\orcidlink{0000-0003-2053-0749}\inst{2,3},
    Andrea Banzatti\orcidlink{0000-0003-4335-0900}\inst{4},
    Enrico Ragusa\orcidlink{0000-0001-5378-7749}\inst{5,6},
    Emma T. Whelan\orcidlink{0000-0002-3741-9353}\inst{7},\\
    Guillaume Bourdarot\inst{8},
    Catherine Dougados\orcidlink{0000-0001-6660-936X}\inst{9},
    Eleonora Fiorellino\inst{10,11}\orcidlink{0000-0002-5261-6216},
    Sean I. Mills\inst{7}
    }

\institute{
    European Southern Observatory, Karl-Schwarzschild-Strasse 2, 85748 Garching bei München, Germany \and
    SETI Institute, Mountain View, CA 94043, USA/NASA Ames Research Center, Moffett Field, CA 94035, USA \and
    Department of Astronomy, The University of Texas at Austin, Austin, TX 78712, USA \and
    Department of Physics, Texas State University, 749 North Comanche Street, San Marcos, TX 78666, USA \and
    Dipartimento di Fisica, Università degli Studi di Milano, Via Celoria 16, 20133 Milano MI, Italy \and 
    Dipartimento di Matematica, Università degli Studi di Milano, Via Saldini 50, 20133, Milano, Italy \and
    Department of Physics, Maynooth University, Maynooth, Co. Kildare, Ireland \and
    Max-Planck-Institut für extraterrestrische Physik, Giessenbachstrasse 1, D-85748 Garching, Germany \and
    Univ. Grenoble Alpes, CNRS, IPAG, 38000 Grenoble, France\and
    Alma Mater Studiorum – Università di Bologna, Dipartimento di Fisica e Astronomia “Augusto Righi”, Via Gobetti 93/2, I-40129, Bologna, Italy \and 
    INAF – Osservatorio Astronomico di Trieste, via Tiepolo 11, I-34143 Trieste, Italy
}

\titlerunning{Monitoring Accretion in DQ Tau}
\authorrunning{H.Alqubelat et al}

   \date{Received September 2025; accepted November 2025}

\abstract{Multiplicity in pre-main-sequence (PMS) systems shapes circumstellar and circumbinary disks, often resulting in morphological features such as inner cavities, spiral arms, and gas streamers that facilitate mass transfer between the disk and stars. Consequently, accretion in eccentric close binaries is highly variable and synchronized with their orbits, producing distinct bursts near periastron passages.
In this study, we examine the orbital and accretion properties of the eccentric Classical T-Tauri binary star DQ Tau using medium- to high-resolution spectroscopy obtained using the Very Large Telescope (VLT) X-Shooter and UVES instruments. The data have been taken at the time of a monitoring of the inner disk chemistry with JWST, and the results of our analysis are needed for a correct interpretation of the JWST data. We refine the orbital parameters of the system and report an increment in the argument of periastron of $\sim 30 ^{\circ}$. This apsidal motion can be caused by the massive disk acting as a third body in the system. We also explore the possibility that the resulting apsidal motion is caused by a still not-detected additional (sub-)stellar companion. In this case, we estimate a lower limit of $\sim 15M_{J}$ for the mass of this putative companion at the cavity edge ($a=3a_{\rm bin}$). We investigate the accretion of the primary and secondary stars in the system using the Ca\,II 849.8 nm emission line. We observe the primary accretes more at the periastron compared to its previous quiescent phases. The secondary dominates the accretion at post-periastron phases. Additionally, we report an elevated $L_{\rm acc}$ at apastron, possibly due to the interaction of the stars with irregularly shaped structures near their closest approach to the circumbinary disk. Finally, we derive the accretion luminosity of each star across the disentangled epochs and compare the results to those derived by the UV excess, finding a good overall agreement. The individual $L_{\rm acc}$ values can be used as an input for the chemical models.
}

\keywords{stars: DQ Tau -- stars: pre-main sequence -- stars: variables: T Tauri -- accretion -- binaries: spectroscopic}

\maketitle

\section{Introduction}

High-angular-resolution observations and theoretical models of young stellar objects (YSOs) have highlighted the impact of multiplicity on the surrounding disk, often manifesting as prominent morphological features such as inner cavities, spiral arms, streamers, and more \citep[e.g.,][]{2023ASPC..534..605B}. Numerous theoretical work put these features in relation to the presence of stellar or planetary companions \citep{2023ASPC..534..423B}. 

Both GG Tau and HD 142527 show strong tidal interactions between their stellar companions and surrounding disks, producing gas flows from circumbinary to circumstellar regions. In both systems, these interactions shape disk structures, including inner cavities, asymmetries, and spiral arms, highlighting the role of companions in driving ongoing disk dynamics \citep{2020A&A...639A..62K, 2024A&A...688A.102T,
2011A&A...528A..91V, 2012ApJ...753L..38B,2024A&A...683A...6N}.

Similarly, AK Sco, with a separation of $\sim 0.16$ au, exhibits periodic enhancements in X-ray and UV emissions, attributed to accretion streams crossing the inner cavity and delivering material onto the stellar surfaces \citep{2013ApJ...766...62G}. Complementary studies of CoKu Tau/4 revealed an inner cavity carved by the binary's gravitational influence \citep{2008ApJ...678L..59I}.

Pre-Main-Sequence (PMS) accretion is a stochastic process giving rise to variability on timescales from hours to decades \citep{hartmann2016}. PMS stars can witness routine-like variability with 1-2 mag change in brightness for up to weeks, while burst-like variability does increase brightness 1-2.5 mag, from a week up to a year \citep{2023ASPC..534..355F}. Meanwhile, the presence of binary companions is expected to produce accretion variability with characteristic frequencies corresponding to their orbital timescales and sub-harmonics, leading to pulsed accretion in eccentric systems \citep{1996ApJ...467L..77A, 2016MNRAS.460.1243R,2020MNRAS.495.3920T}.

Accretion flows in close binary systems are enhanced by dynamical interactions that periodically channel material from the circumbinary disk onto the individual stars. This process tends to preferentially feed the lower-mass companion, driving the system towards mass ratio unity over time \citep{2000A&A...360..997T, 2000AAS...197.1004B}, with differences expected depending on disk parameters, binary mass ratio, and orbital eccentricity, and the precession angle between disk and binary stars\citep{2015MNRAS.447.2907Y}. 
This accretion-driven mass equalisation may be the origin of the observed peak in the mass ratio distribution of close binaries around $q\approx1$.

\citet{2016ApJ...827...43M} performed 2D simulations on accretion flows in the circumbinary cavity of an equal-mass binary. They showed that the circumstellar disks of eccentric binaries are severely truncated at pericenter. The circumstellar disks show differences in their surface densities, implying that the two stars of the equal-mass binary accrete at different rates. This was also shown by \citet{2002A&A...387..550G}, who demonstrated that accretion in close to equal-mass binaries is highly time-variable and not equally distributed among the two components of the system. This effect is driven by asymmetries in the circumbinary flow at the various phases of the orbital period. 

Observational monitoring supports such theoretical models, as accretion variability of young spectroscopic binaries is readily observed. \citet{2017ApJ...842L..12T,2019AJ....158..245T} showed that the eccentric TWA 3A ($e=0.63$, $q=0.84$) experience periodic accretion bursts near periastron passages, with accretion rates increasing by a factor of $\sim4$.  Using the emission line of He\,I $5876\AA$, they showed that circumbinary accretion streams preferentially feed the primary star. In the case of the multiple system VW Cha \citep{2022ApJ...941..177Z}, a peak-to-peak photometric variability amplitude of up to $\sim0.8$ mag was observed, which was attributed to changes in the accretion rate, possibly influenced by the multiplicity and the presence of a circumbinary disk. Similarly, the WX Cha binary system exhibits variability with a photometric amplitude up to $\sim 0.5$ mag \citet{2022ApJ...938...93F}. Another examples is CVSO 104 ($e=0.4$, $q=0.9$), where both components exhibit similar levels of accretion, as indicated by He\,I $6678/5876 \AA$ and Balmer emission line profiles, accreting from a shared circumbinary disk \citep{2021A&A...656A.138F}.

Key questions remain about how binary properties, particularly eccentricity, mass ratio, and orbital phase-modulated accretion dynamics shape the structure of the inner disk. These interactions not only shape disk morphology but are crucial for interpreting time-sensitive spectroscopic observations, especially in systems where both stars are actively accreting, leading to large orbit-to-orbit variability. The impact of variability on the physical and chemical conditions of the inner disk also remains open. The simultaneous accretion monitoring is particularly crucial to be able to interpret the JWST MIR spectra, which are rich in molecular lines, originating from the inner regions of the disk \citep{2023AJ....165...72B,2023ApJ...947L...6G} . Hence, characterising the overall binary accretion behaviour, ideally disentangling the contribution of each component, driven by the binary configuration is essential for understanding the observable chemistry, and structure of the inner disk, where planet formation takes place \citep[Hyden et al. in prep.][]{2025arXiv250819701K}.

This work focuses on DQ Tau, a double-lined spectroscopic binary where both components are of M0 spectral type. 
The system is located at RA 04$^{\mathrm{h}}$ 46$^{\mathrm{m}}$ 53$^{\mathrm{s}}$.058, Dec +17$^\circ$ 00$'$ 00$''$.14, and it has a total mass of $M_{1+2} = 1.21 \pm 0.26\, M_{\odot}$ as derived from radial velocity measurements and ALMA CO disk modeling \citep{2016ApJ...818..156C}.
The orbital period of the binary has previously been reported to be $\sim$15.8 days \citep{Mathieu1997, 2024MNRAS.528.6786P} with the orbit highly eccentric ($e \sim 0.6$). 
This system has become a benchmark for studying pulsed accretion. DQ Tau exhibits recurrent accretion bursts near periastron \citep{2017ApJ...835....8T, 2014ApJ...792...64B,2022ApJ...928...81F,2025arXiv250408029T} with both components actively accreting. Using 8 epochs of spectroscopic data obtained with the ESO Very Large Telescope (VLT) X-Shooter instrument, \citet{2022ApJ...928...81F}  showed that the primary and secondary accretes at different phases in the orbit with the dominant accretor alternating between the two star. Recently, \citet{2024MNRAS.528.6786P}, used the high-resolution CFHT/ESPaDOnS spectrograph to cover one orbital period of this system. They reported that both stars accrete with the primary accreting more compared to previous orbital cycles, resulting in a balanced accretion between the two stars;
While in  \citet{2023MNRAS.518.5072P}, it was noted that the secondary is the dominant accretor. 

The analysis of the VLT/X-Shooter data, to be used also in this work, as well as LCO u$'$ photometry obtained in a joint JWST and ground-based campaign to monitor DQ Tau, revealed that the accretion rate can vary by nearly two orders of magnitude near periastron across ten orbits \citep{2025arXiv250408029T}. In order to correctly interpret the JWST data, a detailed knowledge of the orbital and accretion proprieties of DQ Tau at the time of JWST monitoring is essential.
In this work, we use the ground-based observations (VLT/X-Shooter and VLT/UVES) from this monitoring to refine the orbital parameters of DQ Tau. We then trace the accretion of each star across various orbital cycles to estimate the mass accretion rates of each stellar components, complementing on \citet{2025arXiv250408029T}, who only considered the total accretion rate on the binary.

The paper is structured as follows. A description of the observations is provided in Sect.~\ref{sect:obs}, whereas the
results on the orbital and accretion properties are presented in Sect.~\ref{sect:analysis}. We then discuss the results of the orbital and accretion properties separately and compare them to the literature in Sect.~\ref{sect:discussion}, before concluding this paper in Sect.~\ref{sect:conclusions}.

\begin{figure*}
    \centering
    \includegraphics[width=0.9\textwidth]{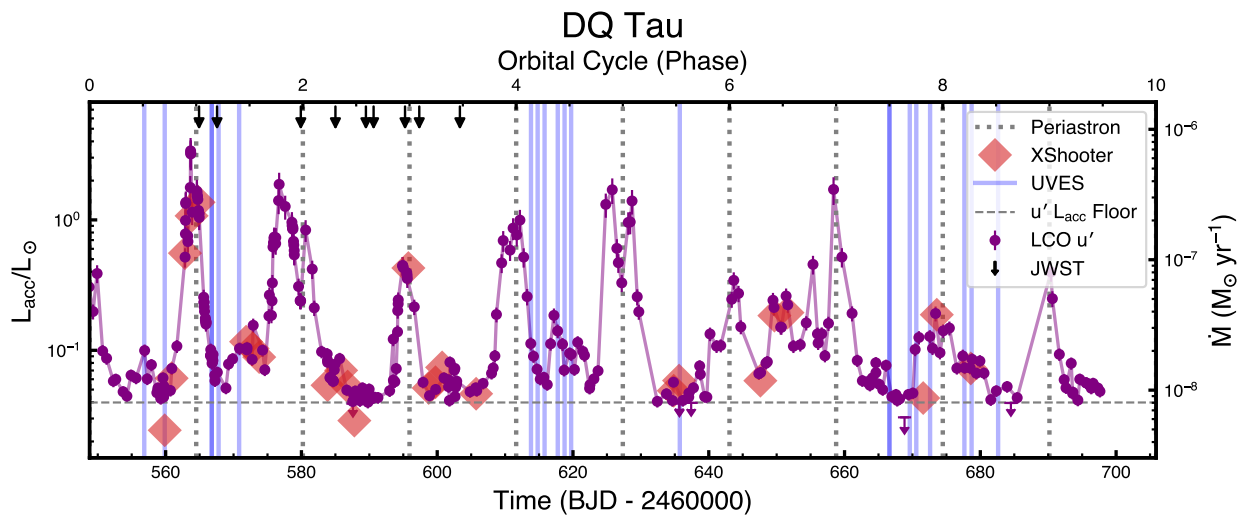}
    
    \caption{Accretion variability of DQ Tau shown by observations with LCO u$'$ photometry, X-Shooter, and UVES taken across 5 months between August 13th, 2024 to January 22nd, 2025. The accretion luminosity (left axis) and mass accretion rate (right axis) are measured from X-Shooter spectra and LCO u$'$-band photometry for the whole system \citep{2025arXiv250408029T}. Time is reported in barycentric Julian days on the bottom axis and the binary orbital cycles are labelled on the top axis. Vertical dotted lines mark the periastron passages of DQ Tau with the JWST MIR observations indicated as black arrows.}
    \label{fig:DQTau_LCO_XS}
\end{figure*}

\section{Observations and data reduction}\label{sect:obs}

In Fig.\ref{fig:DQTau_LCO_XS}, we present an overview of the data sets included in this paper, using the time-series accretion-luminosity measurements for the combined accretion rates measured from \citet{2025arXiv250408029T}.
The binary periastron passages, the time where accretion bursts happen like clock-work, are marked as dotted vertical lines.

\subsection{VLT/X-Shooter}

We observed DQ Tau with the X-Shooter spectrograph \citep{2011A&A...536A.105V} mounted on the ESO Very Large Telescope (VLT) during the program 114.2799.001 (PIs: C.F. Manara, B. Tofflemire). This spectrograph works at medium resolution (R $\sim$10,000 – 20,000) and simultaneously covers the wavelength range $\sim$ 300-2500 nm, dividing the spectra in three arms (UVB, VIS, NIR). Our observations are set to achieve the highest possible resolution, therefore using the narrowest slits (0.5”, 0.4”, 0.4” in the three arms, respectively), while achieving absolute flux calibration accounting for the slit losses with a short exposure with the broad slits of 5.0” width prior to the narrow slit observations. A nodding cycle consisting of four positions ABBA was used for the narrow slit observations to achieve a better sky subtraction at near-infrared wavelengths.The observing strategy was optimized to complement each JWST observation with three X-Shooter spectra. 

An absolute time interval was set to start the X-Shooter monitoring in each flare event covered by JWST–MIRI (see Hyden et al. in prep. for JWST observations strategy), and the subsequent epochs were put in relative time-link intervals with minimum distances between the observations of 10 hours. A total of 25 spectra were taken, 19 of them close in time to the JWST observations between September 5th and November 21st, 2024. Not all of the planned spectra  were taken on the expected time-interval due to different causes, including interventions on the telescopes and visitor mode runs. As a result, a total of six spectra were observed between December 2nd 2024 and January 3rd, 2025 (see Fig.~\ref{fig:DQTau_LCO_XS}). All data were taken with excellent sky transparency conditions (“clear” or “photometric”).

Data were reduced using the ESO pipeline for X-Shooter v.3.6.8 \citep{2010SPIE.7737E..28M} in the Reflex environment \citep{2013A&A...559A..96F}. Telluric lines were removed with the ESO Molecfit tool \citep{2015A&A...576A..77S}, and the final flux calibration was obtained rescaling the narrow slit spectra to the wide slit ones, following the procedure described by \citet{2021A&A...650A.196M}. We estimated the S/N in four nm wavelength window for the visual arm at $\lambda=786$ nm for each epoch. The VIS arm has the highest precision and accuracy of wavelength calibration with a standard deviation of $\sim \ 0.5$ km\,s$^{-1}$. We report the observations log and estimated S/N at wavelength range between $\lambda=784$ nm and $\lambda=788$ nm in Table \ref{table_all}. 

\subsection{VLT/UVES}
During the same monitoring program, DQ Tau was observed also with the UVES spectrograph \citep{2000SPIE.4008..534D} mounted on the VLT during the program 114.27MG.001 (PI: E. T. Whelan).
The motivation for this programme was to use spectro-astrometry to study changes in the spatial properties of the [O I] 6300 emission from DQ Tau (Mills et al. in prep).
We used the RED setting centered at 580 nm, with slit width of $0.8''$, giving a resolution {R $\sim$50,000} and covering the wavelength range $\sim$ 480-680 nm. 
Data were taken between September 2nd, 2024 and January 8th, 2025, always with excellent sky transparency conditions (“clear” or photometric”). A total of 20 epochs were taken (see Fig.~\ref{fig:DQTau_LCO_XS}).\\
Data were reduced using the ESO pipeline for UVES v.6.4.10 \citep{ballester2000} in the Reflex environment \citep{2013A&A...559A..96F}. The data delivered by the pipeline were flux calibrated, although not corrected for slit losses. Data were not telluric corrected. Instead, we avoided regions where telluric lines are strong in the UVES spectrum in the analysis. For each epoch,
we estimated the S/N in four nm wavelength wide region for the red arm at $\lambda=6654$ \text{\AA}.  We report the observations log and the corresponding S/N at wavelength range between $\lambda = 6652$ \text{\AA} and $\lambda = 6656$ \text{\AA} in Table \ref{table_all}.

We also obtained from the ESO archive the data for a template star, Tyc7760283-1, obtained under porgram ID 082.C-0218(A) (P1: C. H. F. Melo). We reduced the spectrum using 
the same version of the ESO pipeline for UVES v.6.4.10 in the Reflex environment as for the DQ Tau spectra.

\section{Analysis}\label{sect:analysis}

In this section, we present the results obtained from the datasets described in Sect.\ref{sect:obs}. We fit the radial velocity (RV) measured on absorption lines to determine the orbital parameters of DQ Tau. Then, we use the Ca\,II $849.8$ nm emission line and Li 670.8 nm absorption line to trace the effect of accretion on line emission and veiling for each component of DQ Tau.

\begin{figure*}
    \centering
    \includegraphics[width=0.9\textwidth]{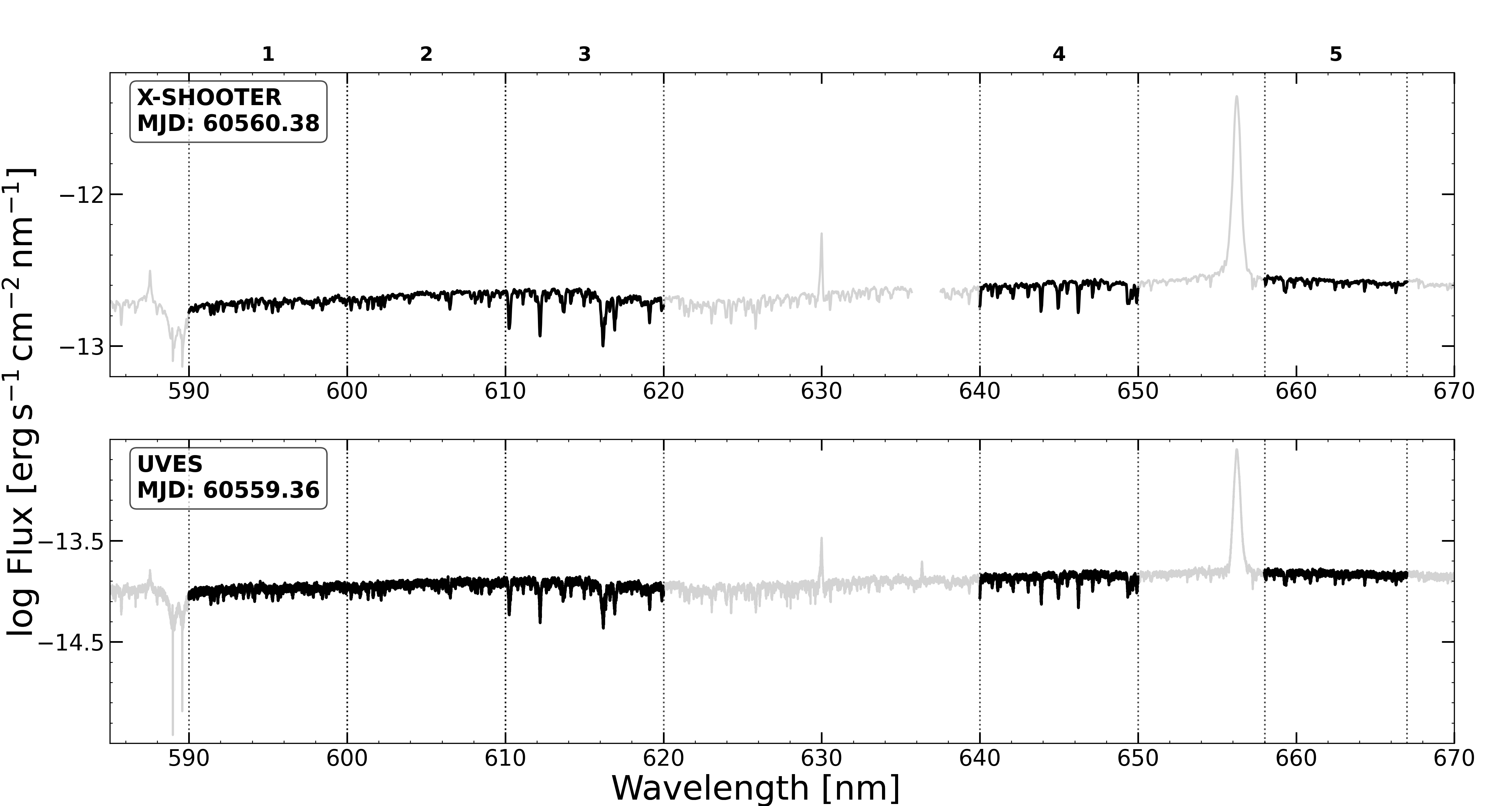} 
    \vspace{3mm}
      \caption{ Spectra of DQ Tau taken with the visual arm of X-Shooter and with UVES. The wavelength ranges in black (590-600, 600-610, 610-620, 640-650, 658-667 nm) correspond to the five regions where the broadening function was calculated for each epoch. The wavelength ranges were chosen to be free of strong emission lines and telluric contamination. } 
    \label{fig:all spec}
\end{figure*}

\subsection{Orbital properties from the RV analysis}
\label{subsection:RVs}

In Fig.~\ref{fig:all spec}, we show examples of X-Shooter and UVES spectra of DQ Tau, covering the wavelength ranges used for the analysis. We select five wavelength ranges in common between X-Shooter and UVES to calculate the RV, each of them ten nm wide. The visual arm spectrum of X-Shooter (450-1000 nm) gives the highest resolution approximately $R\sim 20000$, compared to the UVB and NIR arm, and it is therefore the best suited for this study. For UVES, we use the wavelength range 5800-6800 \text{\AA}, which overlaps with the coverage of X-Shooter, as the rest of the spectrum is affected by the presence of emission lines or strong telluric features. In addition, the bluer parts of the UVES spectrum exhibits lower S/N. 

We use a total of 45 spectra to carry out the radial velocity measurements. Each spectrum is corrected to the barycentric velocity frame using \textit{pyasl.baryCorr}\footnote{\url{https://pyastronomy.readthedocs.io/en/latest/pyaslDoc/aslDoc/baryvel.}}. We then compute the broadening function (BF), implemented in \textsc{SAPHIRES} \footnote{\url{https://github.com/tofflemire/saphires/tree/master}}, for each spectrum using a class III M0 spectral type template star (Tyc7760283-1) with a $v\sin i$ of $14.4 \pm 0.5$  km\,s$^{-1}$ \citep{2013A&A...558A.141S}. We note that this value of $v\sin i$ is in the order of the vsini of DQ Tau \citep{2016ApJ...818..156C}. The X-Shooter spectrum for the template is taken by \citet{Manara13}, while the UVES one is used here for the first time.

The RV of the X-Shooter and UVES templates are obtained using \textsc{ STAR\_MELT}\citep{starcampbell}\footnote{\url{https://github.com/justyncw/STAR_MELT}}, calculating a cross-correlation function (CCF) in $100~\AA$ region, free of emission lines, between $5000 - 6000~\AA$, with the standard RV template star used in the package. 

The final RV of the template is the mean value of all the shifted CCF sections. The RVs of the templates are consistent between the two spectra $\sim \,6.4\, \pm 0.1 ~\mathrm{km\,s^{-1}}$ \,\text{and}\, $6.6\, \pm 0.1~\mathrm{km\,s^{-1}}$
, respectively. The values are in agreement with the RV derived by \citet{2022ApJ...928...81F}. 

The produced BF for each DQ Tau spectrum spans an interval of $-100$ km\,s$^{-1}$
 to $+100$ km\,s$^{-1}$. Double-peaked BFs with clearly separated or blended peaks are fitted with two Gaussian profiles in SAPHIRES. Single narrow-width peaks are fitted with a single Gaussian profile.

The peaks of the BF are considered to be the RV of each component at the time of the observation. The derived RV values are then assigned to the primary and secondary stars based on the fitted amplitudes of each RV curve. By calculating RVs across several wavelength regions, we confirm that the RVs are consistent with each other across the whole range of wavelengths of the spectra for each epoch. Based on this, we calculate the mean RV across the wavelength ranges and the standard deviation for each epoch (Fig.~\ref{fig:bf_wave}).
The extracted mean RV values and errors for each epoch are reported in Table \ref{table_all}.

\begin{figure}[h!]
    \centering
    \raggedright 
   \includegraphics[width=0.5\textwidth]{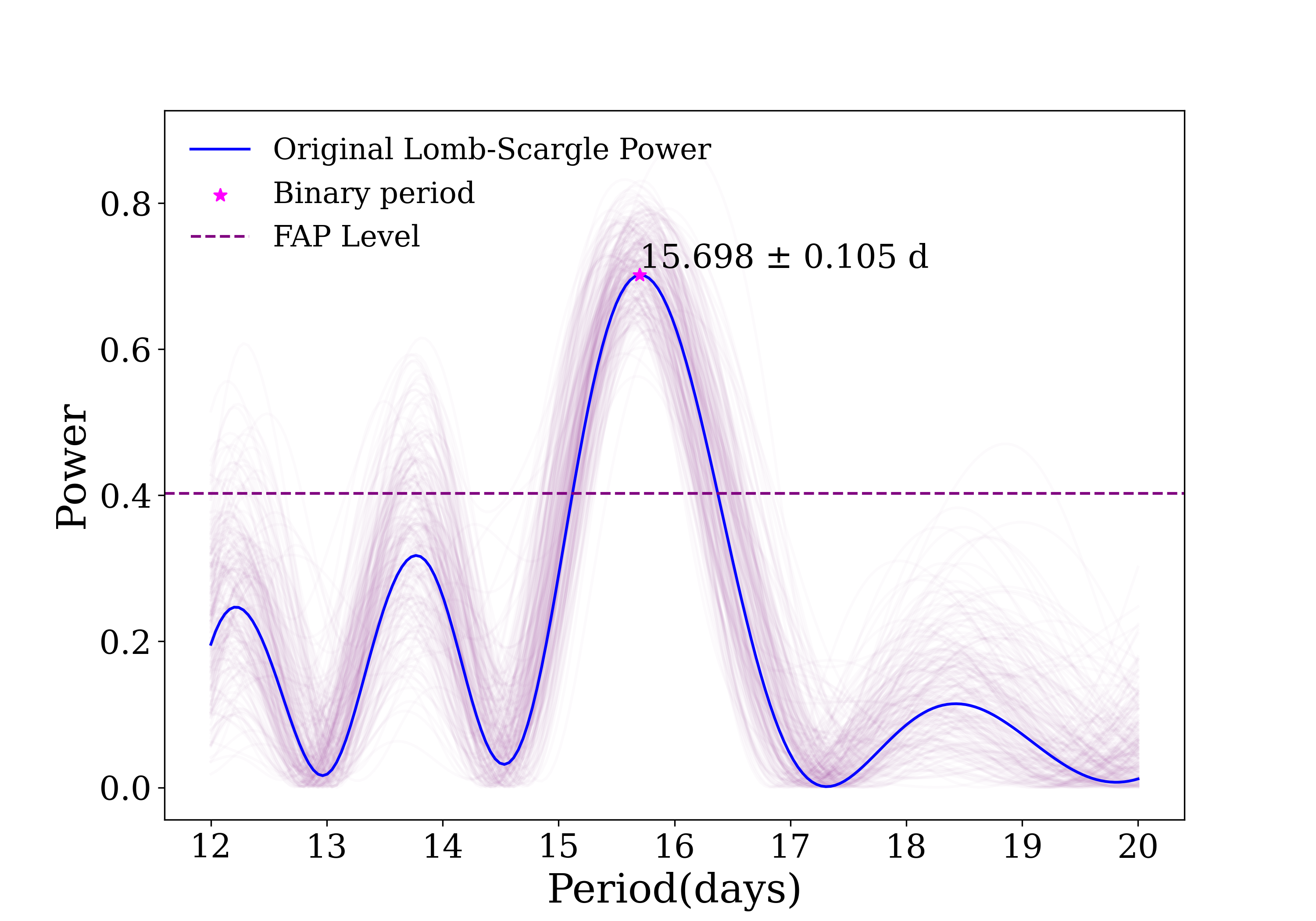} 
    \caption{Lomb-Scargle Periodogram of RV1 data. The data shows a significant peak reported at 15.698 day. The $1\%$ false alarm probability (FAP) level is indicated by the magenta dotted line. In the background, the bootstrapped periodograms are shown in purple.}
    \label{fig:figure_2.png}
\end{figure}

We use a Lomb-Scargle Periodogram to look for the binary orbital period signal on the measured RVs. The periodogram yields a detection at period of $15.698 \pm 0.105$ days when the RVs of the primary or of the secondary are used. A one percent false alarm probability (FAP) threshold is also computed to determine the power level above which signals are considered statistically significant, finding that only this peak is statistically sound (Fig.~\ref{fig:figure_2.png}). To estimate the uncertainty in the detected periodic signal, we employ a bootstrapping approach, generating 200 periodograms by randomly resampling the RV measurements with replacement and calculating the standard deviation of the signal within the 14–16 days range. 

\citet{Mathieu1997} and \citet{2016ApJ...818..156C}, report a period at 15.8 days based on photometric data from Berkeley Automated Imaging Telescope with V, R, and I bandpasses, with a baseline observations of 5000 days. Additionally, \citet{2024MNRAS.528.6786P} reports a period of $P=15.8$
days based on a fit of 11 spectra taken over 40 nights using the Levenberg–Marquardt algorithm. The values of the period derived from our periodogram analysis is thus consistent with the values reported in the literature. Additionally, we use the Markov Chain Monte Carlo (MCMC) approach, described below, to fit for the orbital period along with the other orbital parameters. The derived period from this fit is $P =15.71 \pm 0.07$ days, also consistent with the result from the periodogram analysis and the literature. We thus adopt this latter period estimates in the following analysis.

\begin{table}[h]
    \centering
    \caption{Spectroscopy orbital elements of DQ Tau.
    }
    \label{table:orbital_param}
    \renewcommand{\arraystretch}{1.3} 
    \begin{tabular}{lc}
        \toprule
        \textbf{Element} & \textbf{Value} \\
        \midrule
        P (days) & $ 15.71 \pm $ 0.07\\
        $\gamma$ (km s$^{-1}$) & $ 20.33  \pm 0.07$ \\
        K$_1$ (km s$^{-1}$) & $19.03  \pm $0.14 \\
        K$_2$ (km s$^{-1}$) & $20.10 ± \pm $0.17 \\
        e & $ 0.5463 \pm $ 0.0077 \\
        $\omega$ (degrees)  & $ 264.22 \pm  0.01$ \\
        T$_0$ (MJD) & $60564.50 \pm 0.02 $ \\
        \midrule
        $q = M_2/M_1$ & $1.05622  \pm 0.0095 $ \\
        $M_1 \sin^3 i$ (M$_\odot$) & $0.0573 \pm 0.001$ \\
        $M_2 \sin^3 i$ (M$_\odot$) & $0.0542 \pm 0.001$ \\
        $a_1 \sin i$ (au) & $0.0230 \pm 0.0001$ \\
        $a_2 \sin i$ (au) & $ 0.0243\pm0.0001 $ \\

        $a \sin i$ (au) & $ 0.04733 \pm 0.0002 $ \\

        $a \sin i$ ($R_{\odot}$) & $10.18 \pm 0.04$ \\
        \bottomrule
    \end{tabular}
\end{table}

\begin{figure}[h!]
    \centering
    \raggedright 
    \includegraphics[width=\linewidth]{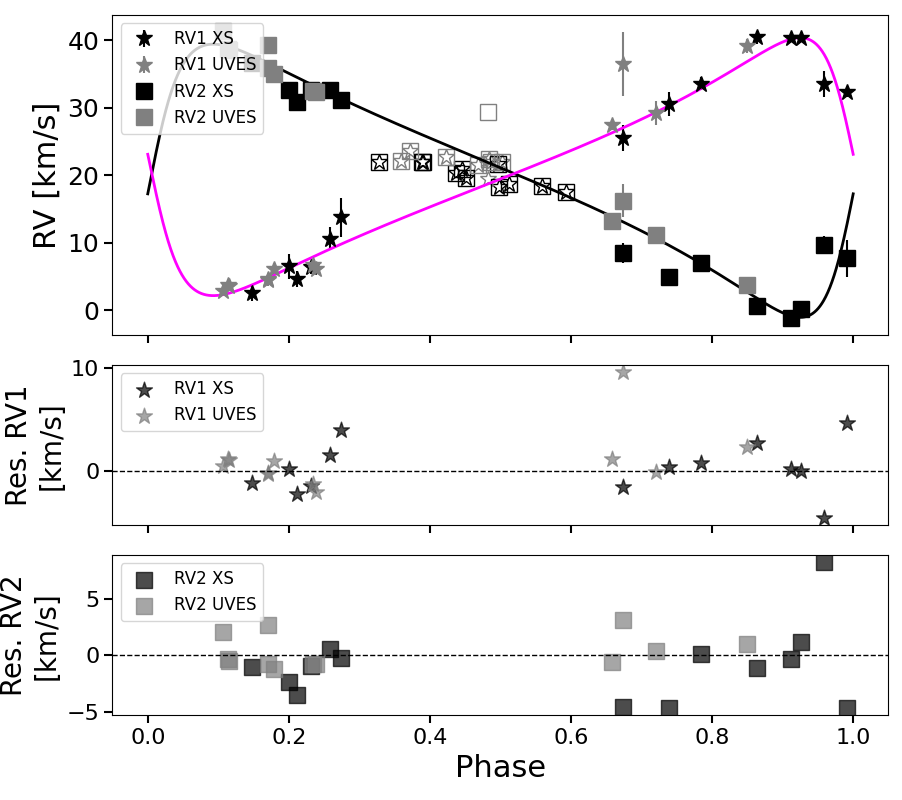} 
    \caption{The orbital solution for DQ Tau in phase. The magenta and black lines fit the primary and secondary data taken with X-Shooter and UVES, respectively. The RVs shown in open symbols at $\sim 0.3-0.6$ in phase are discarded from the fitting procedure. Below, the residuals of the fitted model in phase.}
    \label{fig:combined_phase_fit_res}
\end{figure}

\begin{figure*}[h!]
    \centering
    \begin{subfigure}{0.8\textwidth}
        \centering
        \includegraphics[trim=0 50 0 0,clip,width=\linewidth]{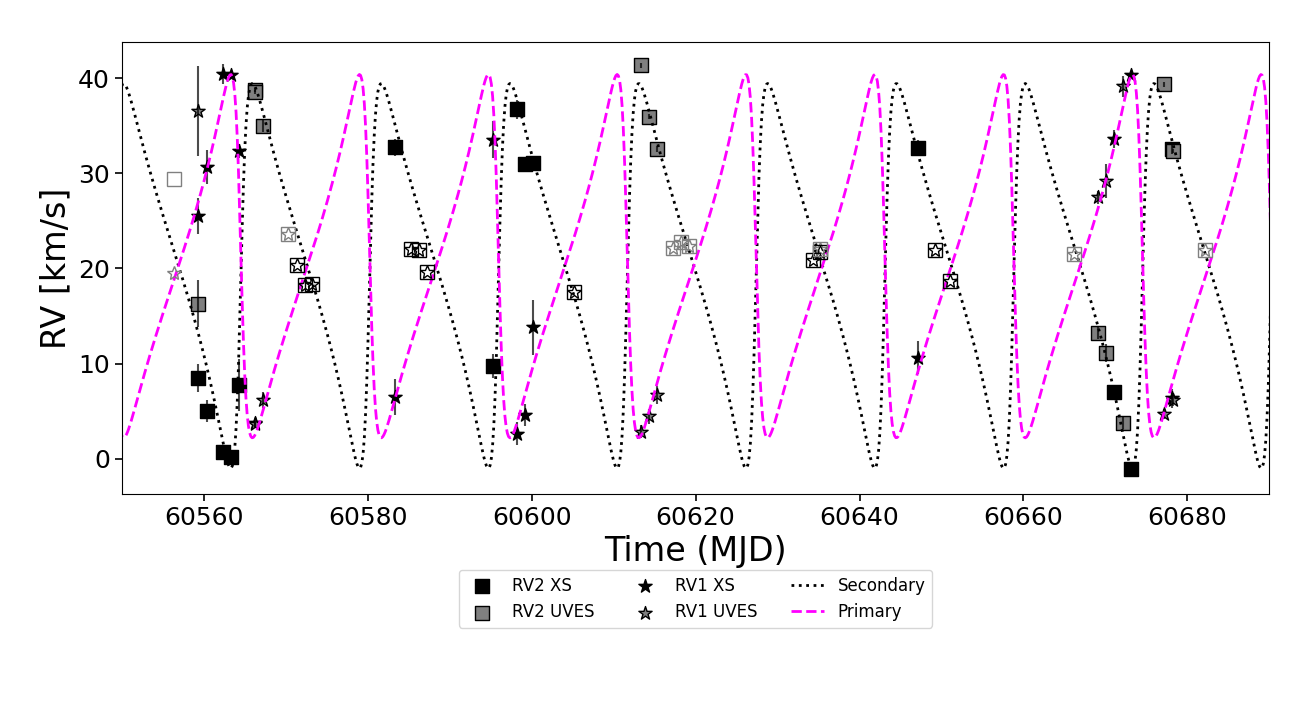} 
    \end{subfigure}   
     \begin{subfigure}{0.72\textwidth}
        \includegraphics[trim=10 0 0 2 , clip,width=\linewidth]{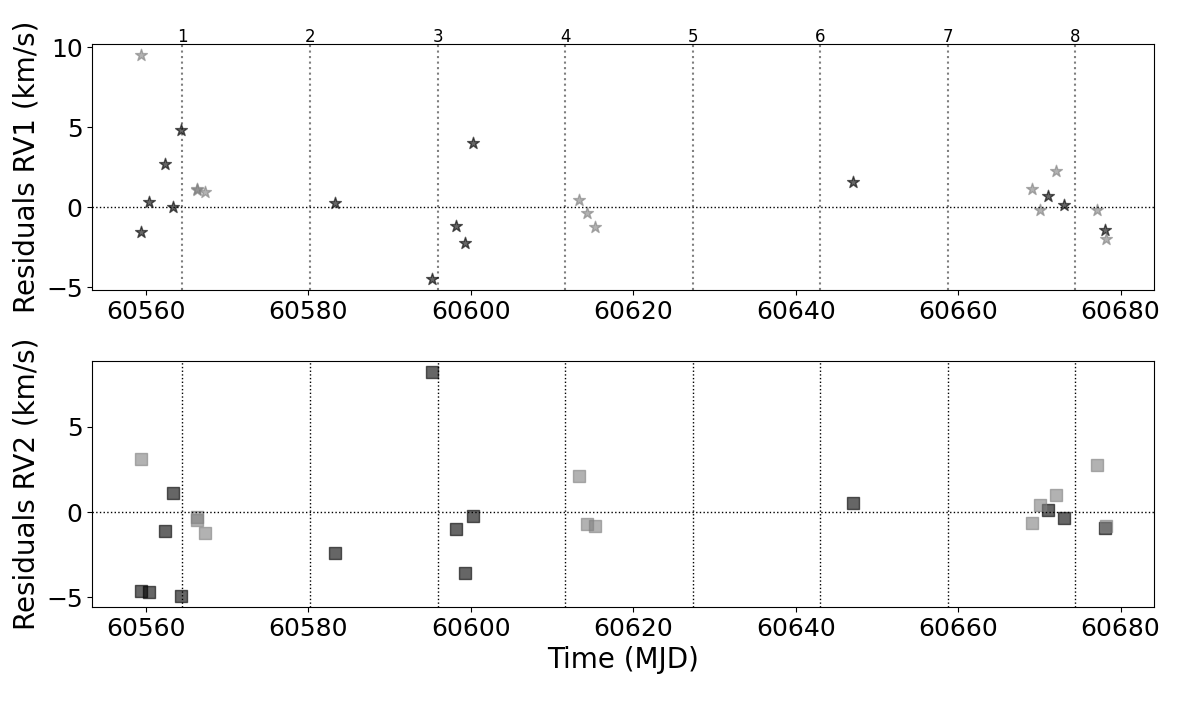}
     \end{subfigure}
  \vspace{-5mm}
    \caption{Orbital solution for DQ Tau across multiple orbits. The magenta and black lines fit the primary and secondary data taken with X-Shooter and UVES, respectively. The RVs shown in open symbols are discarded from the fitting procedure. Below the residual plot. The number of DQ Tau orbital cycles is marked in vertical grey dashed lines.}
    \label{fig:combined_mjd_rv_fit}
\end{figure*}

 We fit the RV data using an MCMC approach to determine the orbital parameters of the binary. The model fits the RV curves of a spectroscopic binary system using a Keplerian orbital model. The radial velocities of the two stars, \( v_1(t) \) and \( v_2(t) \), as functions of time \( t \), derived from the solution to Kepler’s equations. First, the mean anomaly is computed:

\begin{equation}
M(t) = \frac{2\pi (t - T_0)}{P} ~,
\label{eq:mean_anomaly}
\end{equation}
where \( T_0 \) is the time of 
periastron passage and \( P \) is the orbital period. 
This is converted into the eccentric anomaly \( E \) by numerically solving Kepler’s equation: 
\begin{equation}
E - e \sin E = M \,,
\label{eq:keplers_equation}
\end{equation}
with \( e \)  as the eccentricity. The true anomaly \( \theta \) is then calculated from \( E \), and the RVs are given by:
\begin{equation}
v_1(t) =- K_1 \left[\cos(\theta + \omega) + e \cos(\omega)\right] + V_0 \,,
\label{eq:rv1}
\end{equation}
\begin{equation}
v_2(t) = K_2 \left[\cos(\theta + \omega) + e \cos(\omega)\right] + V_0 \,,
\label{eq:rv2}
\end{equation}
where \( K_1 \) and \( K_2 \) are the velocity semi-amplitudes of the primary and secondary stars, \( \omega \) is the argument of periastron, and \( V_0 \) is the systemic velocity.

The model evaluates the log-likelihood by comparing these predicted RVs to the observed RVs of both stars, using observational uncertainties (the standard deviation of the mean RV), calculated across the different wavelength regions, of each spectrum. We use the  \texttt{emcee}\footnote{\url{https://emcee.readthedocs.io/en/stable/} } sampler, with 50 walkers and typical runs of 5000 to 10000 steps. An initial burn-in phase of 500 to 1000 steps is discarded. 
In the fit, we discard the RVs where the BFs from the two stars are blended. The values discarded correspond to epochs at phase ($\phi$) $\sim 0.3-0.6$, shown as open symbols (Fig.~\ref{fig:combined_phase_fit_res}). The resulting fit is consistent with the solution when we consider the whole set of data, with the only exception for the eccentricity ($e$), for which we report a value of $e\sim 0.59$ at $P=15.698$. We use the parameters corresponding to the maximum likelihood values as best estimates of the model parameters. These are reported in the upper part of Table.~\ref{table:orbital_param}.

Fig.~\ref{fig:combined_phase_fit_res} shows the phase-folded fit of the RV data with the residuals. The uncertainty on the parameters is taken to be half the difference between the 84th and 16th percentiles, providing an estimate of the $1\sigma$ confidence interval. We then calculate the projected masses, separations, and the mass ratio.  The corner plot shows the Markov chain resulting from the fit (Fig.~\ref{fig:corner_plot}). 

The MCMC fit is performed leaving the period as a free parameter. On top of that, we also perform a fit using the period obtained from the periodogram, $15.698$ days, and one using the period reported in the literature $\sim 15.8$ days. The orbital solution parameters are consistent in the three cases.

Our data cover a period of more than four months, meaning that we observe almost nine orbits of DQ Tau. In Fig.~\ref{fig:combined_mjd_rv_fit}, we show the RVs reported at different orbital cycles. The RV values discarded from the fit are shown as open symbols correspond to where the broadening function becomes single and therefore is challenging to distinguish the contribution of the primary and secondary at the resolution of X-Shooter and UVES. 

We investigate the potential influence of the accretion variability of DQ Tau on the measured residuals by examining the accretion rates derived by \citet{2025arXiv250408029T}, using X-Shooter and LCO u$'$ photometry. The accretion rates corresponding to the UVES epochs are estimated through linear interpolation of the logarithmic accretion rates, $\log_{10} \dot{M}_{\mathrm{acc}}$, to the respective UVES Barycentric Julian Dates (BJDs). Our analysis reveals no significant correlation between the accretion burst activity near periastron and increases in the radial velocity residuals at any epoch.

\subsection{Accretion properties from lines}
\label{section:accretion}

In this section, we aim to study the accretion properties of each component of the DQ Tau binary system across different orbital phases, to determine which star is dominant in accretion across the nine orbital cycles. 
We first consider the absorption line veiling due to accretion, as measured from the Li 670.8\,nm line as we separate this into two distinct components associated to each star. We discuss in Appendix~\ref{app:veiling} how this information can be used to trace accretion. Then, we aim to use emission lines to measure the accretion rate.
After exploring all the emission lines that trace accretion in the spectra, we notice that most lines are excessively broad or complex (Fig.\ref{fig:HeI_first_cycle}, Fig.\ref{fig:HeI_middle/cycles}, Fig.\ref{fig:HeI_last_2_cycles}), complicating the distinction between the two stellar components contribution, as reported also by \citet{2022ApJ...928...81F}. The best line to be used for the analysis is the bluest of the Ca\,II triplet lines at 849.8\,nm present in the X-Shooter spectra. It has a narrow component and traces the post-shock region close to the stellar surface. The other Ca\,II triplet lines show consistent profiles. In this paper, we use this line to explore the accretion properties of each star.  

\begin{figure}[]
    \centering
    \raggedright 
   \includegraphics[width=0.5\textwidth]{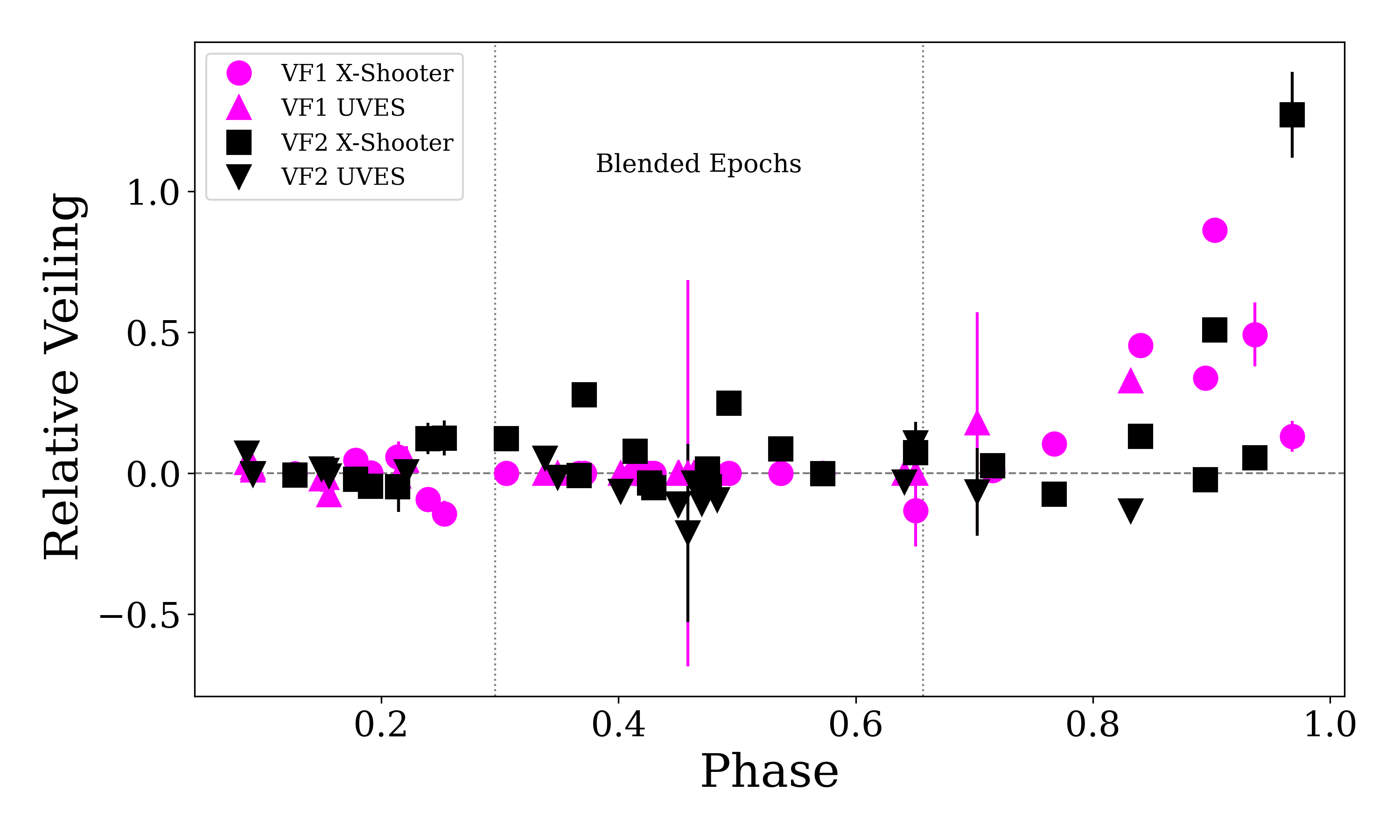} 
    \caption{Relative veiling for primary and secondary using the equivalent width measurements of the Li 670.8\,nm line present in X-shooter and UVES datasets. The dashed vertical lines mark the epochs, where originally one EW value was derived from the Gaussian fitting of line. We then de-blend these values as prescribed in Sec.\ref{section:accretion}. The relative veiling increases in an order of magnitude as the star approaches periastron. }
    \label{fig:veiling.png}
\end{figure}

\begin{figure}[]
    \centering
    \begin{subfigure}[t]{0.49\textwidth}
        \centering
        \includegraphics[width=\textwidth]{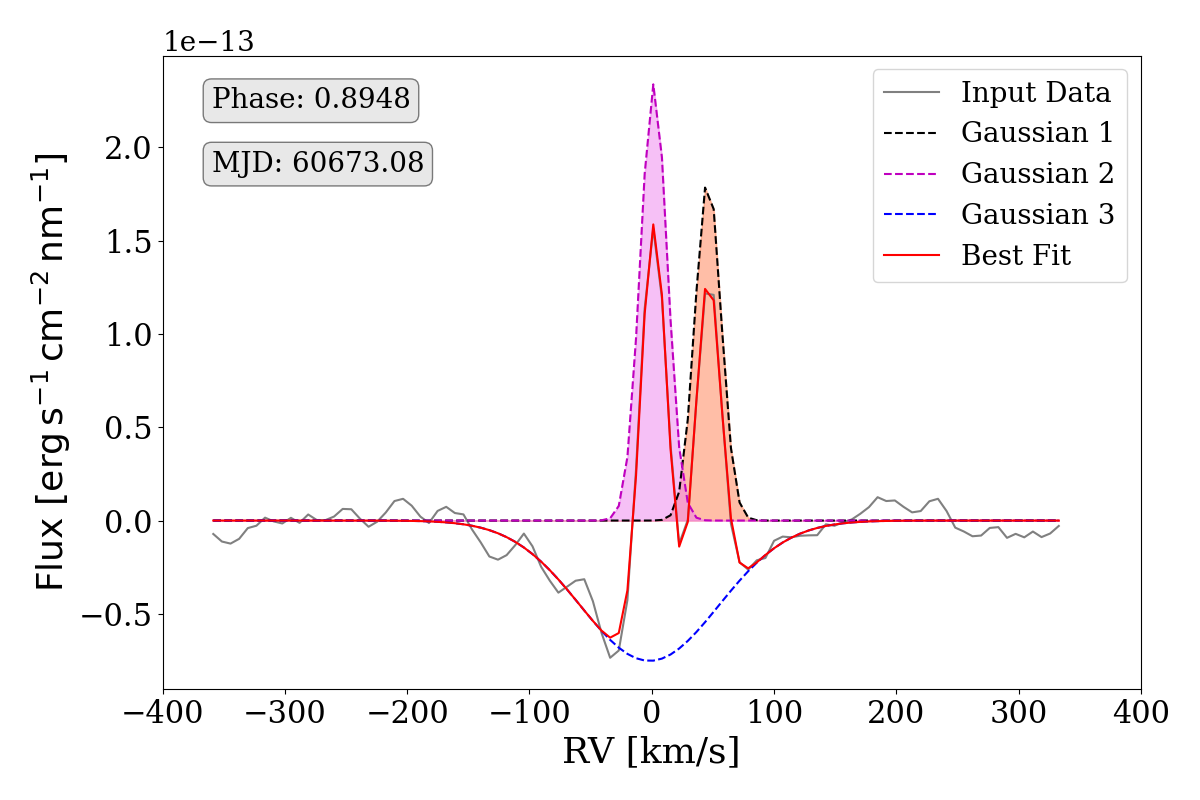}
        \label{fig:gauss_last}
    \end{subfigure}%
    \hfill
    \begin{subfigure}[t]{0.49\textwidth}
        \centering
        \includegraphics[width=\textwidth]{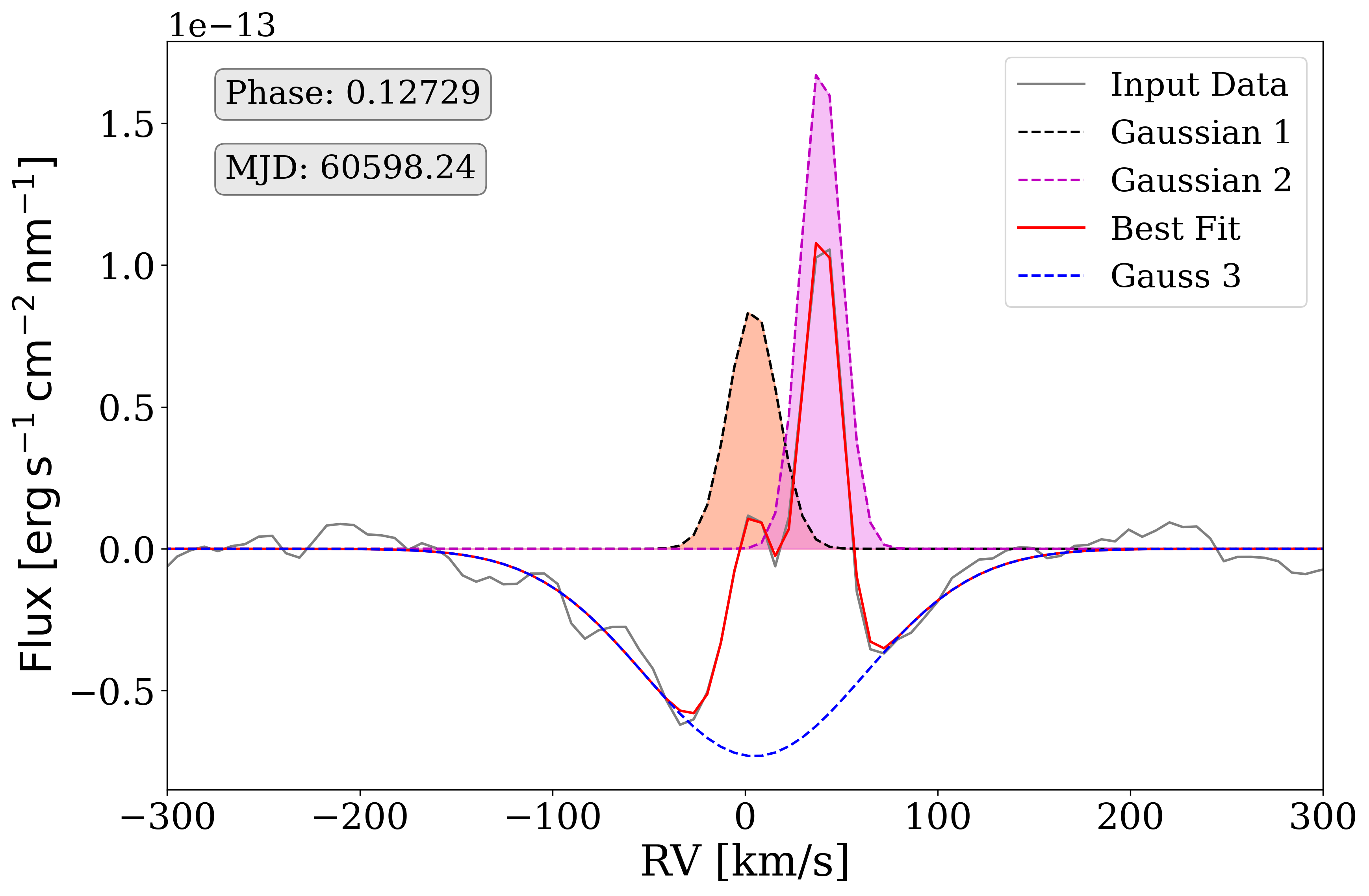}
    \end{subfigure}
    \caption{X-Shooter/Ca\,II 849.8\,nm line at periastron in the last orbital cycle. The double-line feature in grey is fitted using three Gaussian profiles in black, magenta, and blue. The third component in blue is used to subtract the absorption features around each peak in emission. Shaded regions represent the integrated flux of each component. Bottom, X-Shooter/Ca\,II 849.8\,nm in the post-periastron phase of the middle orbital cycles. The double-line feature in grey is fitted using two Gaussian profiles in black and magenta. Shaded regions represent the integrated flux of each component.}
    \label{fig:gauss_fit}
\end{figure}

\begin{figure*}
    \centering
        \centering
        \includegraphics[width=0.9\textwidth, trim=0 30 0 0, clip]{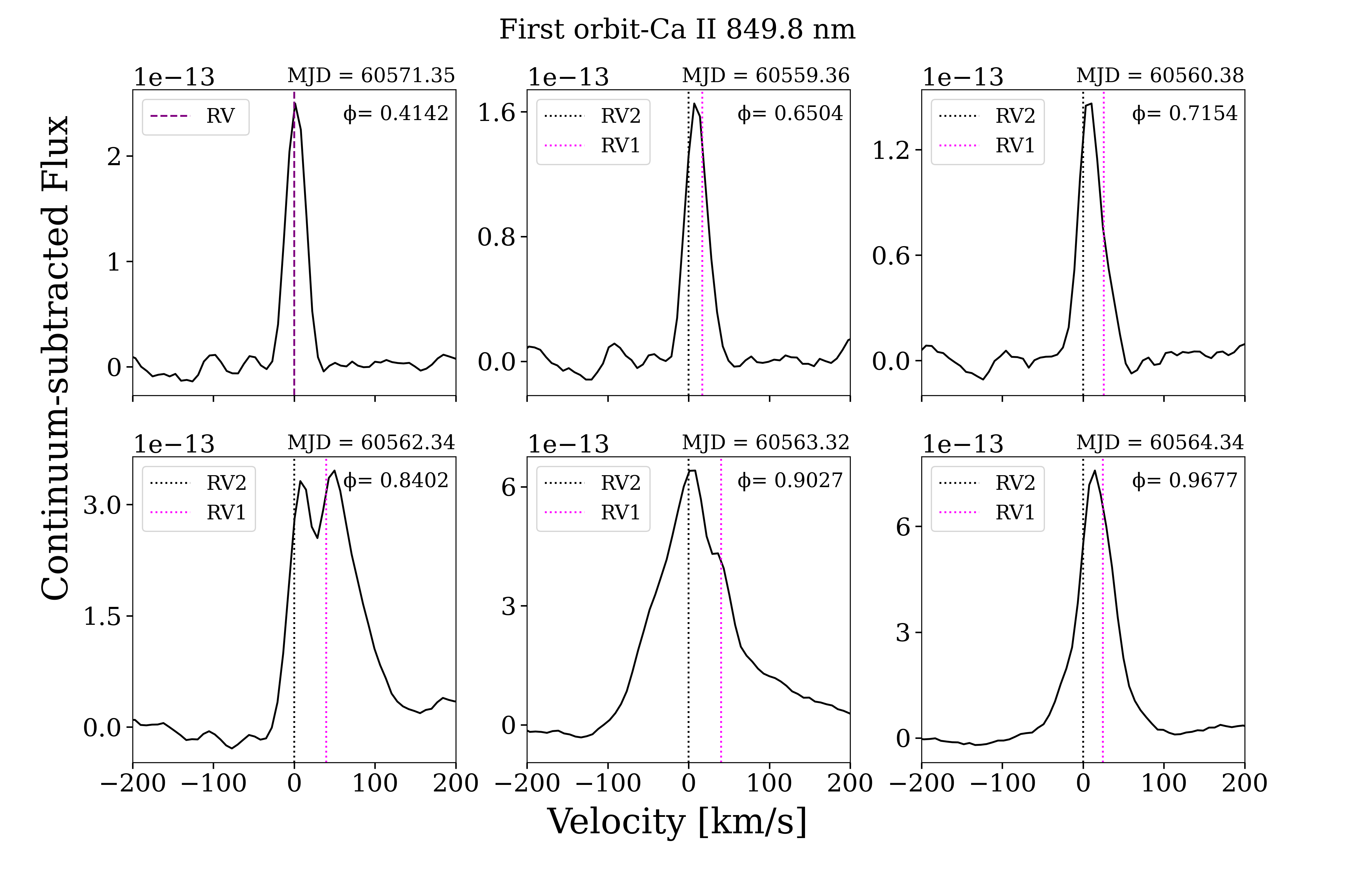}

         \caption{X-Shooter/Ca\,II 849.8\,nm continuum-subtracted line across the first orbital cycle. Orbital phase of each line is indicated in the upper right of each sub-figure. Magenta and black dotted lines show the RVs of the primary and secondary, respectively. Epochs with a single RV value (where $RV_1 = RV_2$) are shown with a purple dotted line. Lines are shifted to the rest-frame of the secondary.}
        
        \label{fig:CaII_first_orb}
        
    \end{figure*}

\subsubsection{Relative veiling measurements}
In order to estimate the veiling on each component, we measure the equivalent width (EW) by fitting two Gaussian profiles on the Li 670.8\,nm line and deriving the EW as:
\begin{equation}
EW_i = \int \left(1 - \frac{F_{\lambda,i}}{F_c} \right) d\lambda ~,
\end{equation}
with $F_{\lambda,i}$ being the best fit profile of each component, labelled with $i$, and $F_c$ the continuum flux. For the fit, we use the information on the estimated RVs of each epoch from Sec.~\ref{sect:analysis} as an initial guess for the absorption peak velocity of each Gaussian component. As shown in \citet{cw_li_23}, we assume that one Gaussian absorption component is sufficient to model the entire contribution of each star to the Lithium absorption line. 

We then estimate the relative veiling for each star in the binary system and at each epoch as 
\begin{equation}
VF_{i}= \overline{EW_{1,2}}/EW_{i}-1\, ,
\end{equation}
where $\overline{EW_{1,2}}$ corresponds to the mean equivalent width for the primary and secondary among the values measured at at $\phi<0.2$. In this analysis, both the X-Shooter and UVES data are used together. At phases $0.3 < \phi < 0.6$ when the two components of the Lithium line from the two stars are fully blended, only one EW value is measured from the data, corresponding to the blend of the two components. 
We infer the $EW_{2}$ of these epochs by subtracting the mean of $EW_{1}$ of the earlier epochs at $\phi<0.2$. We then assume the $EW_{1}$ of these epochs as the mean value. 

We estimate the errors on the EW measurements by generating 50 bootstrapped EW values for each epoch. We introduce random perturbations in the initial guess of the centroid of each Gaussian fitted to the Li 670.8\,nm line from a normal distribution with a standard deviation of 1 km/s. The standard deviation of the measured EW in this bootstrapping is then assumed to be the uncertainty on the EW values. The uncertainty on the epochs were the two lines are blended are assumed to be twice the initial error on the blended EW value. We then propagate these uncertainties to the relative veiling measurement.\\ 

In Fig.~\ref{fig:veiling.png}, we show the relative veiling estimated for each epoch across the phase. 
The relative veiling increases as the star approaches periastron $0.7<\phi<1.0$, and reaches more than unit value as the accretion bursts arise (Fig.~\ref{fig:DQTau_LCO_XS}). The  measured relative veiling of the primary ($VF_{1}$) shows how the accretion drastically increases by $\sim 1$ dex at periastron compared to its previous quiescent state at preceded phases.


\begin{figure*}
    \centering
\includegraphics[trim=0 80 0 50 , clip,width=\textwidth]{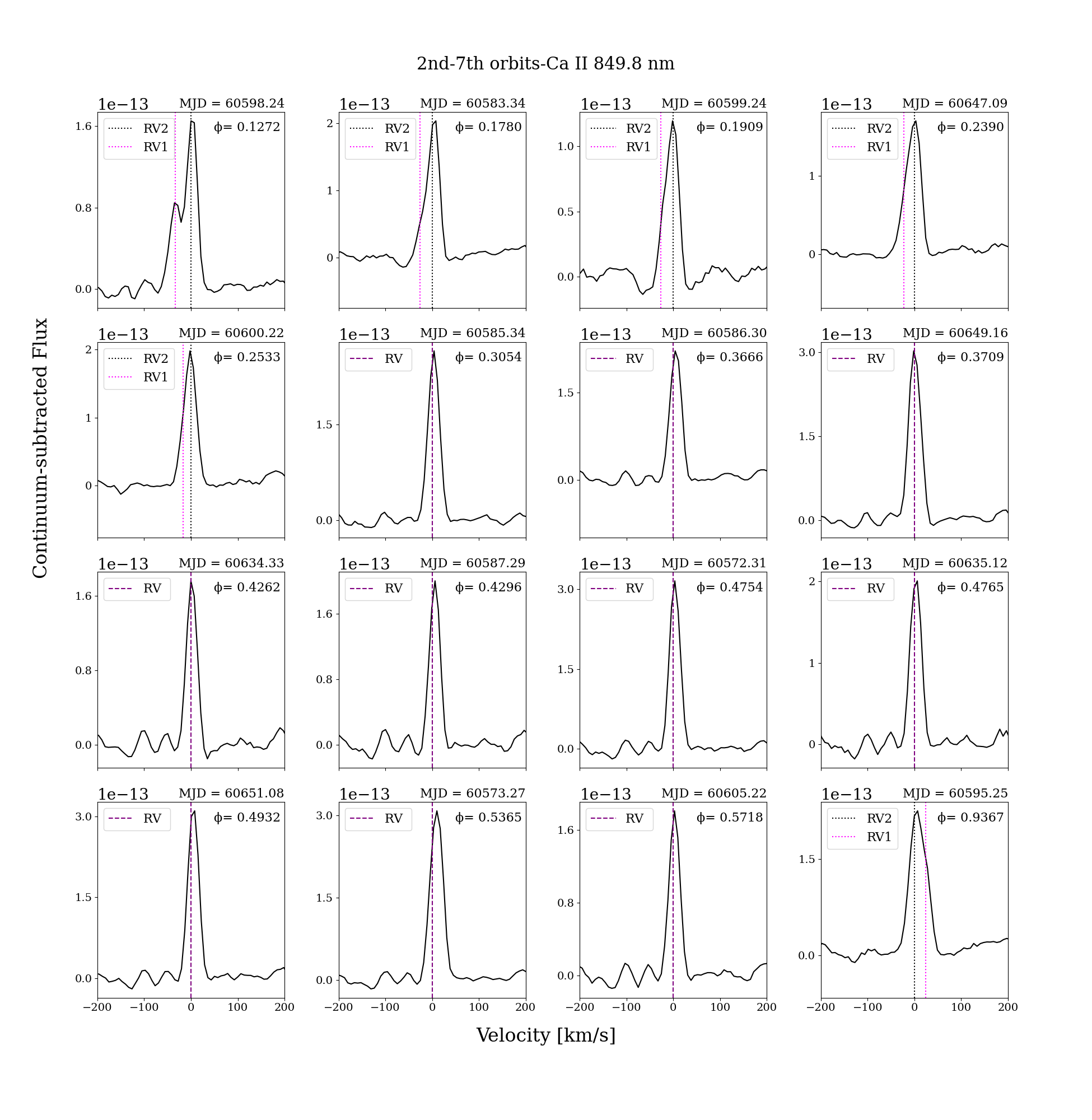} 
      \caption{X-Shooter/Ca\,II 849.8\,nm continuum-subtracted line across the second to seventh orbits. Orbital phase of each line in indicated in the right upper part of each sub-figure. Magenta and black dotted lines indicate the RVs of primary and secondary, respectively. Epochs with One RV value correspond to epochs with single BF, where $RV_{1}=RV_{2}$ are indicated with purple dotted line. The lines were shifted to the rest-frame of the secondary. }
\label{fig:CaII_middle_orb}
\end{figure*}
    
    \begin{figure}
        \centering
        \includegraphics[width=0.5\textwidth, trim=10 17 0 0, clip]{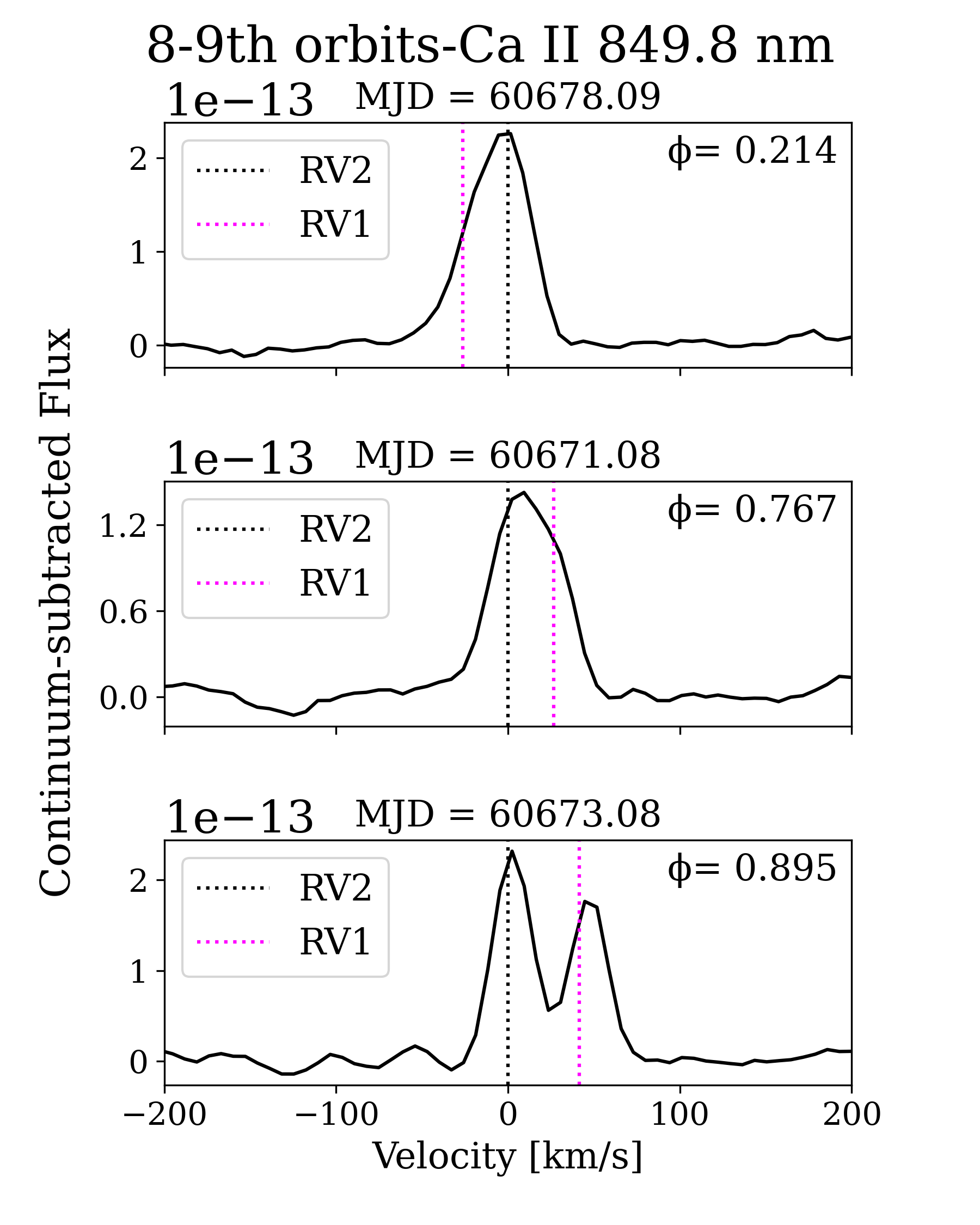}
    
    \caption{X-Shooter/Ca\,II 849.8\,nm continuum-subtracted line across the last orbital cycles. Orbital phase of each line is indicated in the upper right of each sub-figure. Magenta and black dotted lines show the RVs of the primary and secondary, respectively. Epochs with a single RV value (where $RV_1 = RV_2$) are shown with a purple dotted line. Lines are shifted to the rest-frame of the secondary.}
    \label{fig:CaII_last_orb}
\end{figure}

\subsubsection{Calcium emission line}
We perform the removal of the photospheric absorption line and continuum subtraction locally around the Ca\,II 849.8\,nm line by fitting each epoch with two or three Gaussian profiles and a linear component for the continuum. The second or third Gaussian component is used to account for the photospheric absorption around the emission line, when it is present, which must be removed to better trace the accretion of each star (Fig.\ref{fig:gauss_fit}). We then measure the flux of each Gaussian component readily from the continuum-subtracted flux-calibrated spectra.

In Fig.~\ref{fig:CaII_first_orb}, Fig.~\ref{fig:CaII_middle_orb}, and ~\ref{fig:CaII_last_orb}, we show the residual Ca\,II 849.8\,nm line in all the X-Shooter spectra in different orbits, centering the velocity scale on the secondary component. 
We infer that the primary/secondary is accreting the most when the highest peak velocity corresponds to the RV of one of them. 

During the first orbital cycle (Fig.~\ref{fig:CaII_first_orb}), DQ Tau accretes at all phases, including apastron ($\phi=0.414$), but as it approaches the periastron $<0.8402<\phi<0.9679$, we observe the line intensity becomes stronger as both stars are accreting with the secondary growing in flux, shown by the double-line feature of Ca\,II 849.8\,nm line. However, the intensity of the primary emission line peak does not increase at the last two phases in the orbit compared to the secondary.

In the second to the seventh orbits (Fig.~\ref{fig:CaII_middle_orb}), we see that the secondary line intensity is stronger, possibly because it accretes more at $0.1274<\phi<0.2533$. While at $0.3055<\phi<0.5722$, we observe a single line with one RV tracing both components, i.e, hard to attribute the accretion to any of the stars. At the periastron, i.e, $\phi<0.9367$, we observe that the two stars line intensities are lower with respect to the previous periastron phases.
Finally, in the last two orbital cycles (Fig.~\ref{fig:CaII_last_orb}), the Ca\,II 849.8\,nm line has the same line intensity approximately at and post periastron phases, indicating similar accretion rate. 

The variability of the Ca\,II 849.8\,nm double-line feature across the different orbits corresponds to the accretion variability of DQ Tau as indicated in the Fig.~\ref{fig:DQTau_LCO_XS}. The accretion of both stars is more strongly defined in the first orbital cycle.

\begin{figure}[]
    \centering
    \raggedright 
   \includegraphics[width=0.5\textwidth]{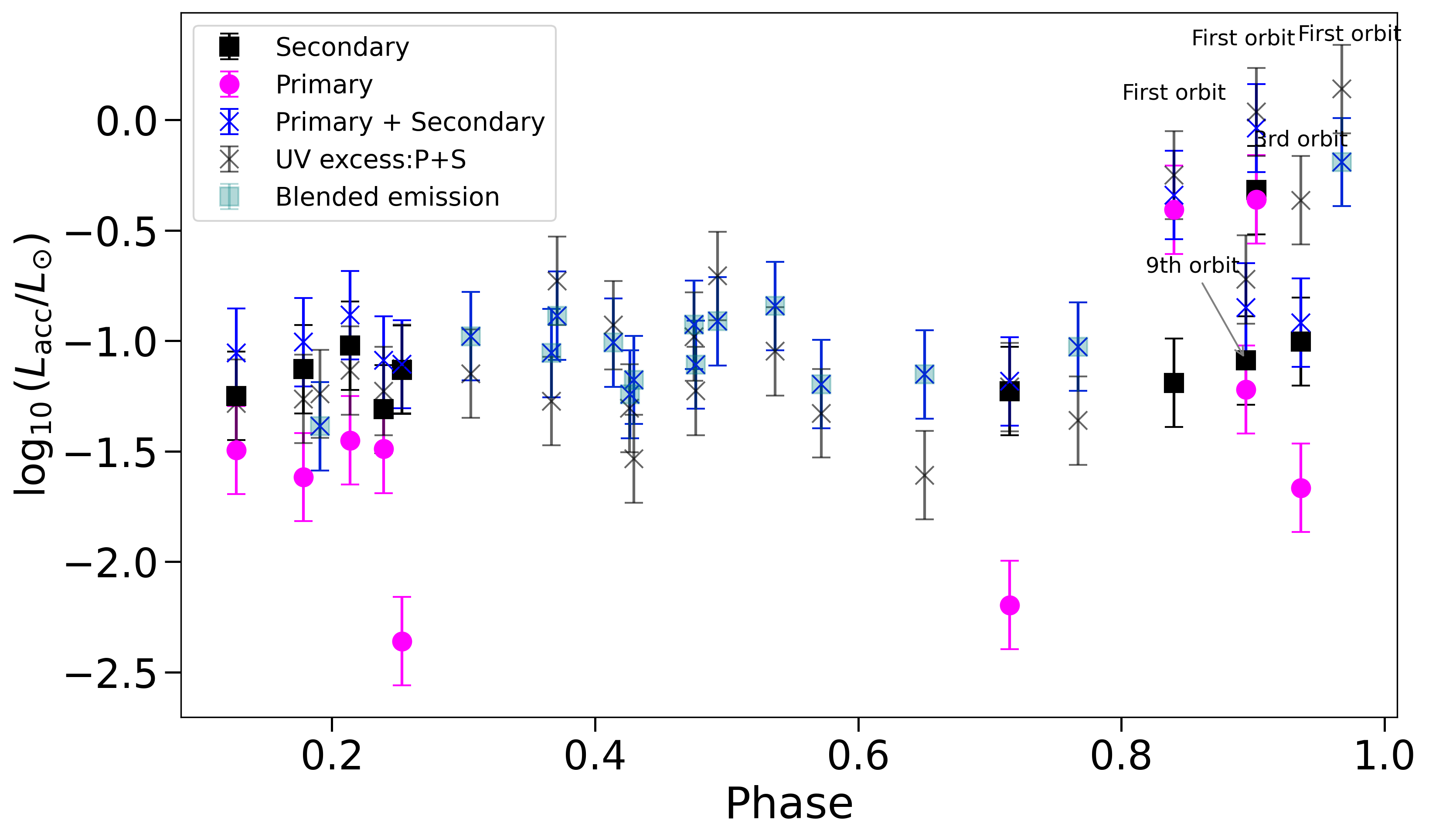} 
    \caption{X-Shooter/Ca\,II 849.8\,nm accretion luminosity measured for each star in DQ Tau in comparison to \citet{2025arXiv250408029T} for DQ Tau as a single star. The sum of the total accretion luminosities for each component are in line with the values derived by \citet{2025arXiv250408029T}. Blended epochs where we derive one integrated flux value are shown as light blue squares. The Secondary appears to dominate the accretion at the disentangled epochs. As DQ Tau approaches periastron, $L_{\rm acc}$ increases. }
    \label{fig:accretion_comp.png}
\end{figure}

To derive the accretion luminosity ($L_{\rm acc}$) of each star, we use the calculated integrated flux for the Ca\,II 849.8\,nm line at different phases for the primary and the secondary. This can only be done at the phases, where it is possible to distinguish the emission peaks of both stars. 
The derived flux across the orbital phase show how variable and modulated is the accretion of DQ Tau and that the secondary is contributing more to the measured flux of the Ca\,II 849.8\,nm line. The flux ratio is elevated during the first orbital cycles at periastron, indicating that both stars are actively accreting. Different periastron passages bring variable flux ratios, confirming the variation of the strength of the accretion bursts as the epochs of the third and ninth orbits show lower values as shown in Fig.~\ref{fig:DQTau_LCO_XS} and Fig.\ref{fig:accretion_comp.png}.

We derive $L_{\rm acc}$ for each star using the integrated flux values on the Ca\,II 849.8\,nm line. We correct the derived flux for extinction at each epoch assuming an extinction value $A_V=1.6$ mag and the Cardelli extinction law \citep{1989ApJ...345..245C}. The estimated reddening factor at $\lambda_{\,\rm Ca\,II\,849.8\, \rm nm}$ is $\sim 0.45$. We first estimate the line luminosities $L_{\rm line,i}$ for each component scaling the flux for the distance as $L_{\rm line,i} = 4 \pi d^{2} F_{\rm line, i}$, where \( d \) is assumed to be 196\,pc \citep{2023A&A...674A...1G}. The accretion luminosity is then derived using the relation: 
\begin{equation}
\log \left( L_{\mathrm{acc},i} \right) = a \cdot \log \left( \frac{L_{\rm line,i}}{L_\odot} \right) + b ~,
\end{equation}

where the values for the $a$ and $b$ coefficients are 1.11$\pm0.13$ and 3.71$\pm0.47$, as derived by \citet{2025arXiv250921078F}.  We assume an uncertainty of 0.2 dex on the disentangled $L_{\rm acc}$. We note that the relations used here are calibrated using rmission line fluxes not corrected for the chromospheric contribution. Therefore, we do not correct for the chromospheric emission , which will be a considerable source of flux at the lowest accretion value

In Fig.\ref{fig:accretion_comp.png}, we compare the accretion luminosity derived from the integrated flux of Ca\,II\,849.8\,nm line to the values estimated by \citet{2025arXiv250408029T} from the UV-excess. We note that the latter assumed that the UV-excess traces the total accretion on the system. 
When the contribution to $L_{\rm acc}$ of each component is summed, the total value is in line with the values derived by \citet{2025arXiv250408029T} from the UV-excess. 
We disentangle the accretion of the components at periastron $0.7<\phi<0.9$, and post-periastron phases $0.1<\phi<0.25$. At other phases, we derive one value of $L_{\rm acc}$ as the two components are fully blended. 

The overall estimated $L_{\rm acc}$ follows the trend shown by \citet{2025arXiv250408029T}, where DQ Tau appears in quiescent accretion mode until it approaches periastron. During the post periastron phases, the derived total luminosity values are in the same order of the epochs taken at apastron. The accretion luminosity values of each component across the phase show that the secondary is accreting at higher pace at post-periastron phases with evident difference in accretion near periastron. At periastron, different epochs from different DQ Tau orbital cycles show variable accretion of both stars. As shown by the Ca\,II 849.8\,nm line peak emission, the primary accretes more during the burst in the first orbit. The periastron phase during the third and ninth orbits show that the secondary is the dominant accretor during these bursts.

In Fig.\ref{fig:lacc_veiling_relation}, we show that the relative veiling values for both components are tracing the accretion. 
The integrated flux, equivalent width, relative veiling, and $L_{\rm acc}$ values for all epochs are provided in Table \ref{table_flux_ratios}.

\section{Discussion}\label{sect:discussion}
\subsection{Apsidal motion}

The orbital properties of DQ Tau derived here, in particular the $e\sim0.55$ and mass ratio $q\approx 0.97$, are in line with previously published papers on DQ Tau (e.g, \citealt{2024MNRAS.528.6786P} and \citealt{2016ApJ...818..156C}).
  
 We report a prograde apsidal motion compared to \citet{2016ApJ...818..156C}. 
 We fit $ \omega=263.1996 \degree \pm   0.0081$, instead of $231.9 \degree \pm 1.8$ \citep{2016ApJ...818..156C}. We emphasize that this precession differs from that of the circumbinary disk cavity induced by the binary’s quadrupole potential, which cannot be detected with present observations in DQ Tau but is nevertheless expected to occur on theoretical grounds \citep{2020MNRAS.499.3362R, 2024MNRAS.532.3166P}.
 
 The precession of the binary orbit could be explained by the gravitational influence of the disk on the binary parameters. \citet{2024ApJ...964...46T} in their hydrodynamical simulation, showed that an accreting eccentric binary experiences prograde apsidal precession faster than changes in the semi-major axis or eccentricity. Hence, the orientation of periastron shifts over time, which also affects the accretion behaviour across the orbit, periodically changing which component of the binary is the dominant accretor and causing significant variations in their individual accretion rates \citep{2015MNRAS.448.3545D}. Across 10 years, the resulted increment in the argument of periastron is $\Delta \omega \sim 30^\circ$. This means DQ Tau would complete one full precession cycle, with approximately 2770 orbits, over 120 years.
 As a first simplified approach to investigate the origin of the observed precession, we estimate the mass that a putative third body in the system would need to reproduce it.  
 We use the approximation by \citet {1999ssd..book.....M} to estimate the mass of this putative companion. First, we define the apsidal precession rate as :

\begin{equation}
\label{eq:dotvarpi}
\dot{\omega}
\simeq
\frac{3}{4}\,\frac{M_{\mathrm{out}}}{M_{\mathrm{bin}}}
\left(\frac{a_{\mathrm{bin}}}{a_{\mathrm{out}}}\right)^{\!3}\,\Omega_{\mathrm{bin}} ~,
\end{equation}
where $M_{\rm out}$ is the mass of the external companion, $M_{\rm bin}$ is the binary mass, $a_{\rm  bin}$ is  the binary semi-major axis, $a_{\rm out}$ is the semi-major axis of the external companion, and $\Omega_{\rm bin}$ is the binary mean motion.
Then, we consider the binary precession timescale compared to the binary orbital scale :

\begin{equation}
\frac{t_{\mathrm{prec}}}{t_{\mathrm{bin}}}
=
\frac{\Omega_{\mathrm{bin}}}{\dot{\omega}}
=
\frac{4}{3}\left(\frac{a_{\mathrm{out}}}{a_{\mathrm{bin}}}\right)^{\!3}\frac{M_{\mathrm{bin}}}{M_{\mathrm{out}}} ~,
\label{eq:9}
\end{equation}

\begin{figure}[]
    \centering
    \raggedright 
   \includegraphics[width=0.5\textwidth]{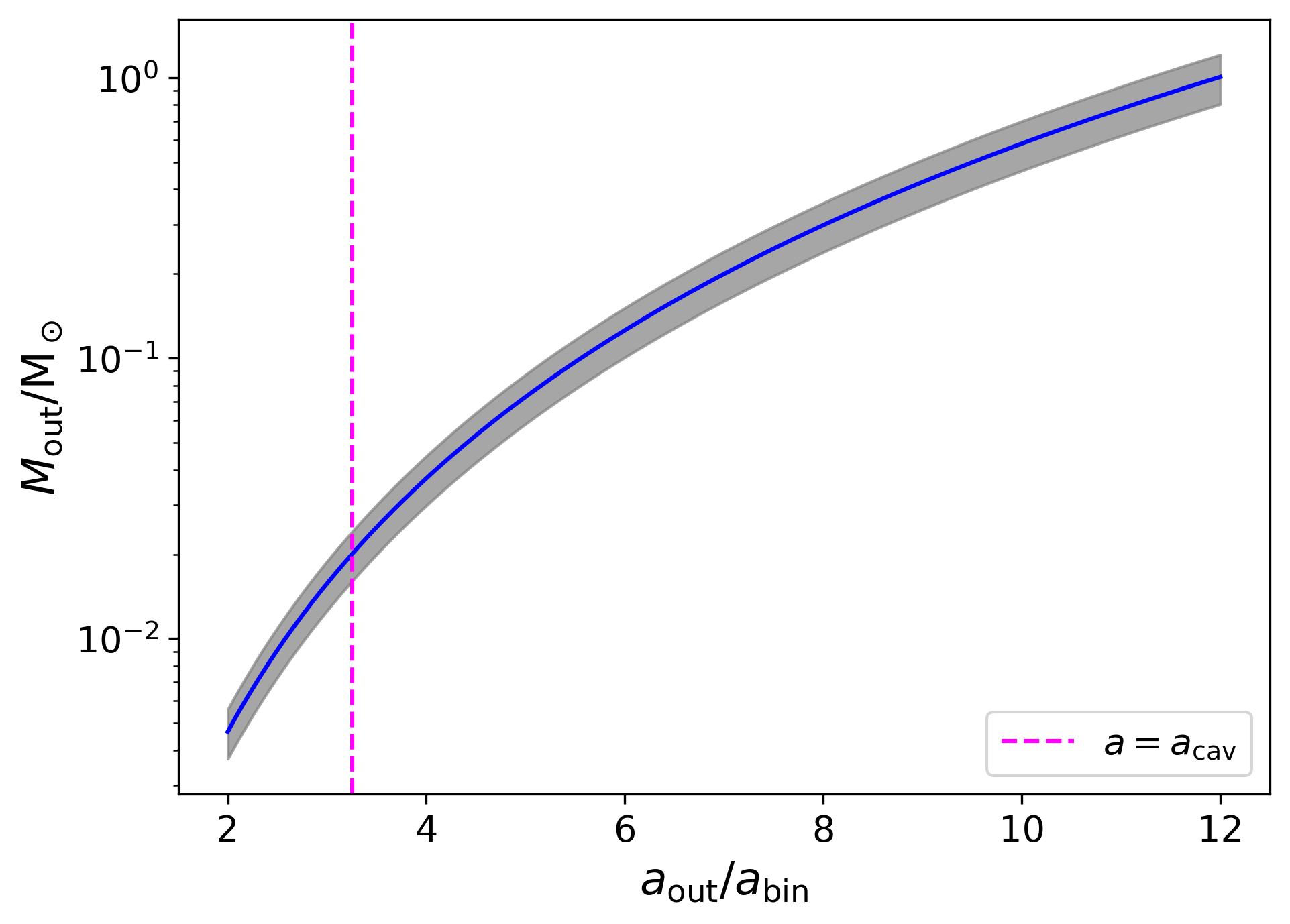} 
    \caption{Mass of a putative third companion as a function of its orbital separation relative to the binary,  required to reproduce the observed apsidal precession of DQ Tau. The solid blue curve shows the corresponding mass estimate, the grey shaded area indicates the uncertainty range, and the vertical magenta dashed line marks the predicted cavity radius($\sim 3a_{\rm bin}$), based on DQ Tau orbital properties and assuming a circular coplanar disk \citep{2025A&A...698A.102R}.}
    \label{fig:third_companion}
\end{figure}

Eq. \ref{eq:9} reduces to the following :

\begin{equation}\label{eq:Mout}
M_{\mathrm{out}}
=
\frac{4}{3}\left(\frac{a_{\mathrm{out}}}{a_{\mathrm{bin}}}\right)^{\!3}
\frac{M_{\mathrm{bin}}}{\,t_{\mathrm{prec}}/t_{\mathrm{bin}}\,} ~\textbf{ ,}
\end{equation}
which describes the relation between the mass and semi-major axis that a putative companion should have in order to explain the observed precession rate.
For DQ Tau, we estimate the mass range of a putative companion responsible for the observed precession rate between $15M_{J}$ at the cavity edge ($\sim 3a_{\rm bin}$) and 1 $M_{\odot}$ at a separation of $\sim 12a_{\rm bin}$ (Fig.\ref{fig:third_companion}). The latter would have been detected, if existed, as it would disturb the disk and carve a cavity. The dust disk mass of  DQ Tau is derived to be $75 M_\oplus$ \citep {2023ASPC..534..539M} based on the ALMA data by \citet{Long19}. Assuming a gas-to-dust ratio of 100, the disk mass would be $\sim25 M_J$, resulting in a ratio of $\sim 0.02$ compared to its total $M_{\rm bin} \sim 1.21M_{\odot}$.  This implies that the precession timescale of DQ Tau could be explained by the disk as shown by \citet{2018MNRAS.474.4460R}, where they modelled the effect of the disk on the precession of the inner companion as an effective additional planet in the system.\\

\subsection{Accretion rate modulation}

Using the Ca\,II 849.8\,nm line and the relative line veiling due to  spots and accretion, as measured from the Li 670.8\,nm line, we provided detailed phase-resolved characterisations of accretion onto the individual components in DQ Tau. While the primary star appears to accrete more at the periastron phases, the secondary star dominates at post periastron passage. Both stars strongly accrete near the periastron in the first orbit. The derived $L_{\rm acc}$ of each component shows that the accretion onto the primary star is  $\sim 1.5$ dex higher at periastron compared to the previous quiescent state phases.

 In our analysis, we observe DQ Tau accreting at all phases, including at apastron. 
Many previous works
\citep{2017ApJ...835....8T,2023MNRAS.518.5072P,2024MNRAS.528.6786P} also observed accretion events at apastron. \citet{2014ApJ...792...64B} observed an anomalous flare at apastron ($\phi = 0.372$ and $\phi =0.433$) with a mass accretion rate that is an order of magnitude stronger than the quiescent rate. This was interpreted due to perturbations in the circumbinary disk, leading to the formation of irregularly shaped, non-axisymmetric structures extending inward from the inner edge of the disk as shown by simulations of \citet{2002A&A...387..550G}. The presence of such structures coinciding with the close passage of one or both stars at apastron, may have led to that unique accretion flare. In our data, we observe an elevated $L_{\rm acc}$ at apastron as the binary is closest to the cavity edge and strip more material.

Our results are in line with to \citet{2022ApJ...928...81F} , in which they covered two epochs of each of the four orbital cycles studied. They reported that the primary was the dominant accretor at the periastron passage and at others was the secondary. The main accretor in their data also changed depending on which accretion line tracer was considered. This phase-dependent switching occurred because different lines originate in different regions in the accretion columns such that new material was accreting onto one star while older material is still accreting onto the other.

\citet{2023MNRAS.518.5072P} used different accretion tracers like $H\alpha$, $H\beta$, $H\gamma$, Ca\,II and the narrow component of HeI $5876 \AA$ line, and reported that the secondary component was the main accretor. However, in a subsequent work \citep{2024MNRAS.528.6786P}, they concluded that the accretion behavior of the DQ Tau components seemed to be balanced.  This was interpreted as the result of the evolution of the primary large-scale magnetic field. In both papers, ESPaDOnS data were taken with time gap of $\sim 2$ years. This alternating accretion behavior was also observed by \citet{2022ApJ...928...81F}, using data taken between 2012-2013. Building on these results, our observing campaign shows that the accretion behavior of DQ Tau is very complex. The primary  and the secondary components are actively accreting near the several periastron passages, while the secondary star is the strongest accretor during the other phases. The several orbits coverage illustrates that the variability in which component dominates in accretion strength and in the overall accretion intensity is extremely time-dependent, confirming the necessity to monitor for longer time close binary systems like DQ Tau to better constrain models of accretion in binary systems. Furthermore, in equal-mass binary star systems as DQ Tau, accretion is characterised by the variability of material arriving through the tidal streams from the edge of the cavity. 

In summary, tracing the accretion pattern of DQ Tau over the past decade has shown a fluctuating pattern between periods of balanced accretion and times when either the primary or secondary star dominates the accretion process. Building upon previous research, our multi-orbit campaign demonstrates that DQ Tau accretion behavior is highly variable over time, with the intensity of accretion activity and the leading accretor changing from one orbit to another. This accretion variability changes the inner disk properties, in particular its temperature, which affect the chemistry and the process of forming planets, as shown by \citep{2025arXiv250912898P}. The availability of this information on the binary interaction is of paramount importance to interpret the JWST molecular emission data obtained for DQ Tau.

\section{Conclusions}
\label{sect:conclusions}
In this work we calculated the radial velocity of the two components of DQ Tau across four months, about 10 orbits, using X-Shooter and UVES spectra. We derived the orbital parameters of the system. Using the integrated flux of the Ca\,II\,849.8 nm, we measured the accretion luminosity on each component across the orbital phases. 

We conclude the following:
\begin{itemize}
    \item The orbital solution we provide is in agreement with previously published results, i.e., \citet{2024MNRAS.528.6786P}.
    \item We report prograde apsidal motion, which could be caused by the disk, acting as a third body in the system.
    \item Alternatively, we explore the influence of a third body interacting with the binary and causing the observed precession. Such object should have a mass of at least $\sim 15 M_{J}$ at a separation of $\sim 3a_{\rm bin}$.
    \item By analysing the accretion lines using only the Ca II triplet, we were able to trace the accretion of both stars. 
    We confirm that the primary and secondary stars accrete differently across the phase and that they alternate in the accretion dominance at periastron. 
    \item The summed values derived from disentangling accretion on each component are in line with the global values measured by \citet{2025arXiv250408029T} using the UV-excess on the unresolved spectra of DQ Tau.
\end{itemize}

The values of the orbit and of the accretion rates derived in this work should be used for the analysis of the JWST spectra of DQ Tau obtained at the time of the observations described here, as well as those obtained shortly after \citep{2025arXiv250819701K}. A modeling of the properties of the inner disk of this complex system should be carried out considering both the orbital dynamics and the accretion rate variability derived in this work.
Similar coordinated ground- and space-based observing campaigns should be set up for future studies of time-variable inner disk chemical properties.

\begin{acknowledgements}
We thank the anonymous referee for their useful report. We thank Myriam Benisty, Cathie Clarke, and Jeff Bary for insightful discussions. Funded by the European Union under the European Union’s Horizon Europe Research \& Innovation Programme 101039452 (WANDA). EF has received funding from the European Research Council (ERC) via the ERC Synergy Grant ECOGAL (grant 855130). Views and opinions expressed are, however, those of the author(s) only and do not necessarily reflect those of the European Union or the European Research Council. Neither the European Union nor the granting authority can be held responsible for them. ER acknowledges financial support from the European Union's Horizon Europe research and innovation programme under the Marie Sk\l{}odowska-Curie grant agreement No. 101102964 (ORBIT-D), including a secondment carried out at the ESO Headquarters in Garching, Germany.

\end{acknowledgements}

\bibliography{aa57425-25}

\begin{appendix}

\section{RV fit and broadening functions}

\begin{figure}[h!]
\sidecaption
    \centering
    \includegraphics[width=0.5\textwidth]{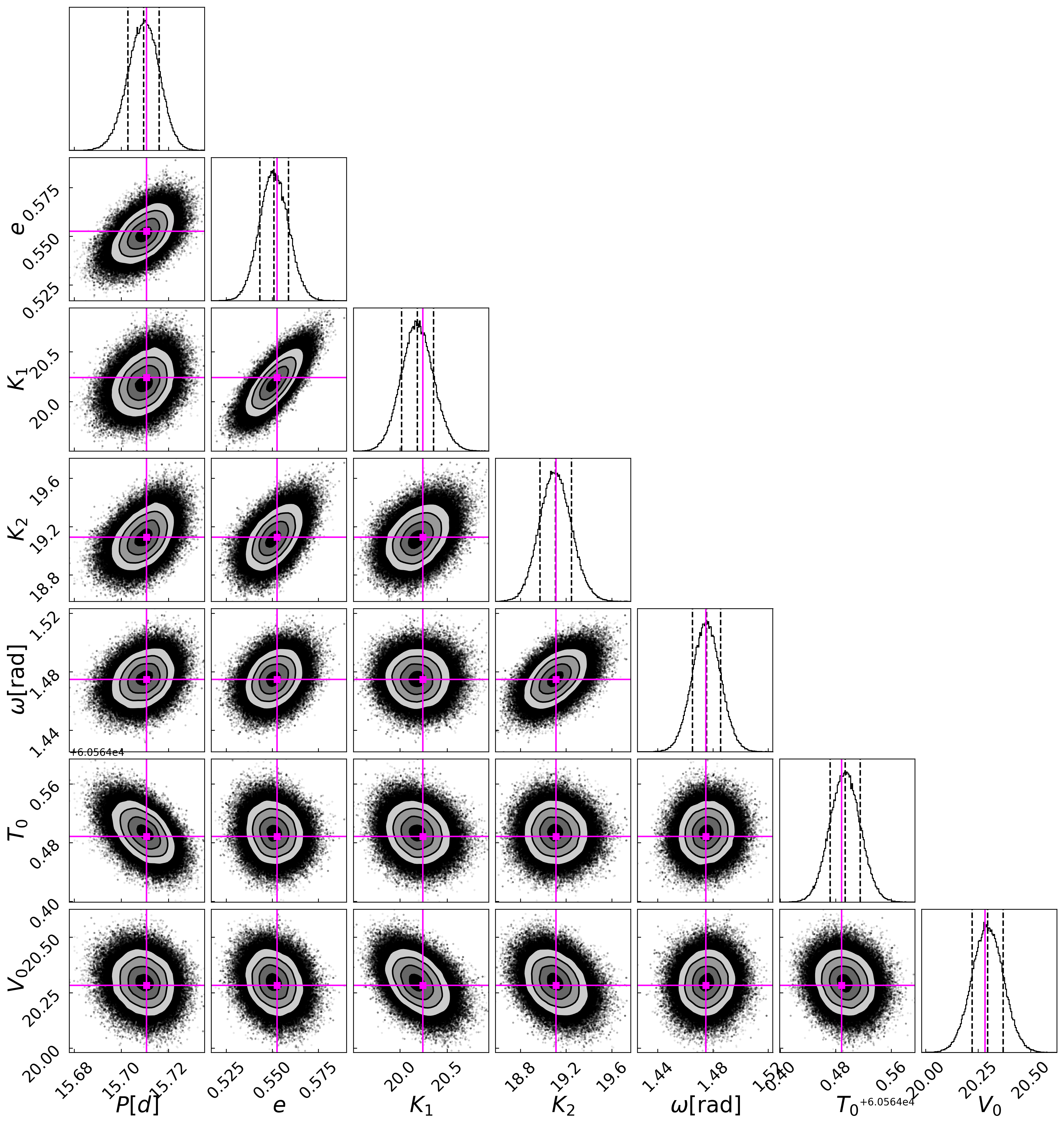} 
    \caption{ Corner plot of the orbital solution for DQ Tau. The vertical dashed lines indicating the 16th, 50th and 84th percentile. The vertical magenta lines correspond to the best likelihood estimates of each distribution.}
    \label{fig:corner_plot}
\end{figure}

\begin{figure}
\sidecaption
    \centering
    \includegraphics[width=0.6\textwidth]{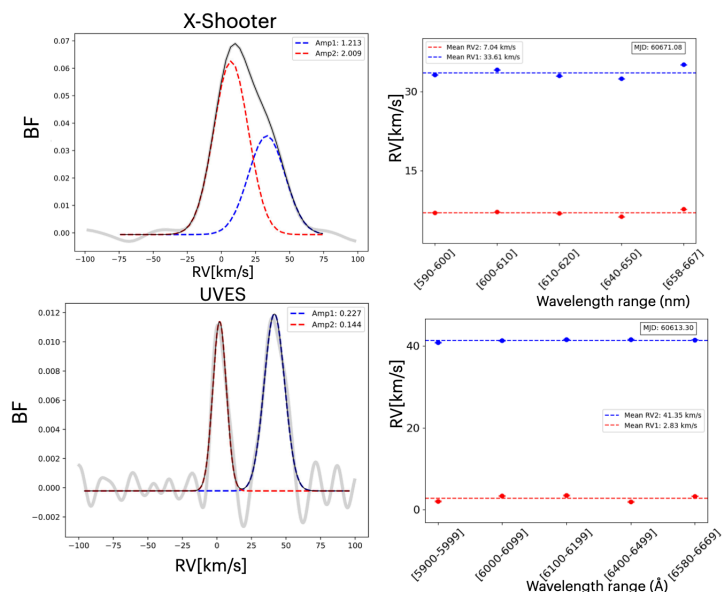} 
    \caption{X-Shooter and UVES broadening functions. Right, the calculated RVs across one spectrum for each spectrograph. The dotted lines represent the mean values.}
    \label{fig:bf_wave}
\end{figure}

The fit of the RV curve described in Sect.~\ref{subsection:RVs} is performed with an MCMC tool, which allows one to explore the correlation between the posterior distribution of the fitted values. This is shown in Fig.~\ref{fig:corner_plot}. 
The distributions are single-peaked, with Gaussian profiles. The magenta lines in Fig.~\ref{fig:corner_plot} highlight the location of the best likelihood estimates. The off-diagonal 2D plots represent the bivariate distributions between each pair of parameters which provide an immediate estimate of their correlation. Additional plots of the BF and of the dependence of RV with wavelength (Fig.~\ref{fig:bf_wave}) are shown here.

\section{Changes in relative veiling and relation with accretion}\label{app:veiling}
 
Measuring the veiling of each individual component of a spectroscopic binary system is not trivial \citep[e.g.][]{2019AJ....158..245T}. Indeed, in a combined light spectrum the additional continuum due to accretion makes the absorption lines shallower with respect to the combined continuum. This additional continuum, however,  affects absorption lines from both the primary and secondary equally, regardless of which star it is coming from. The variations observed here in the EW values could thus be caused by both veiling and/or by spots \citep[e.g.][]{2014ApJ...792...64B} or rotational modulation. 

\begin{figure}[h!]
    \centering
    \includegraphics[width=0.5\textwidth]{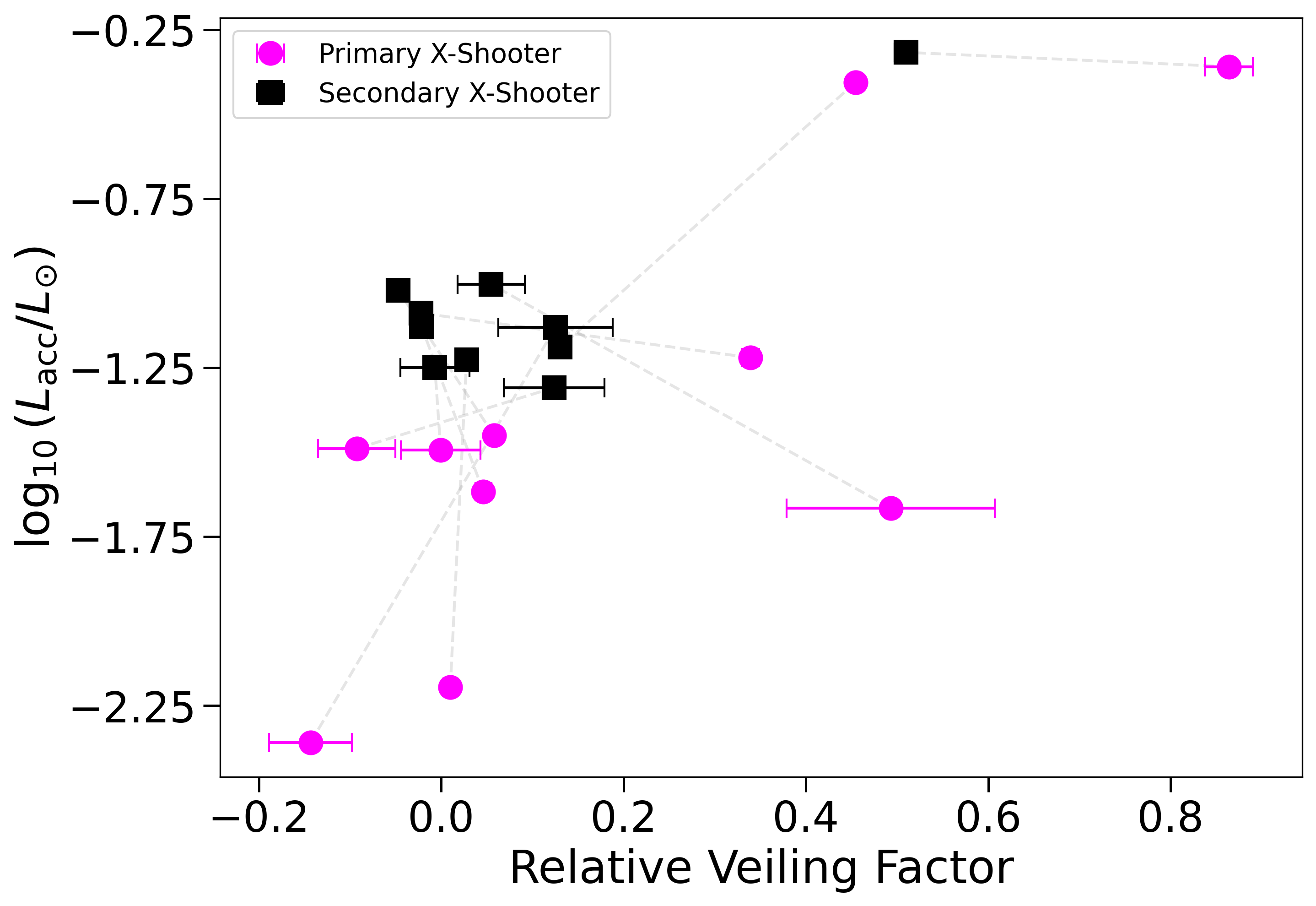} 
    \caption{Accretion luminosity $L_{\rm acc}$ of the primary and secondary stars as a function of the relative veiling. The dotted lines connects the two stars of each epoch, allowing direct comparison of their accretion activity. The data follows a trend, where higher veiling factors are generally associated with larger $L_{\rm acc}$.}
    \label{fig:lacc_veiling_relation}
\end{figure}

In order to test whether the EW variations we see are related to accretion, we plot in Fig.~\ref{fig:lacc_veiling_relation} the measured $L_{\mathrm{acc}}$ for the primary and secondary stars, derived from the Ca\,II 849.8 nm line, as a function of the relative veiling factor measured from the Li 670.8 nm line. 
We see that the observed variations of veiling in both components of the line suggest that the relative veiling variations are mainly tracing the accretion variability in DQ Tau.

\clearpage

\section{X-Shooter and UVES/HeI 587.56 emission line}

\begin{figure}[h!]
    \centering
    \includegraphics[width=0.6\textwidth]{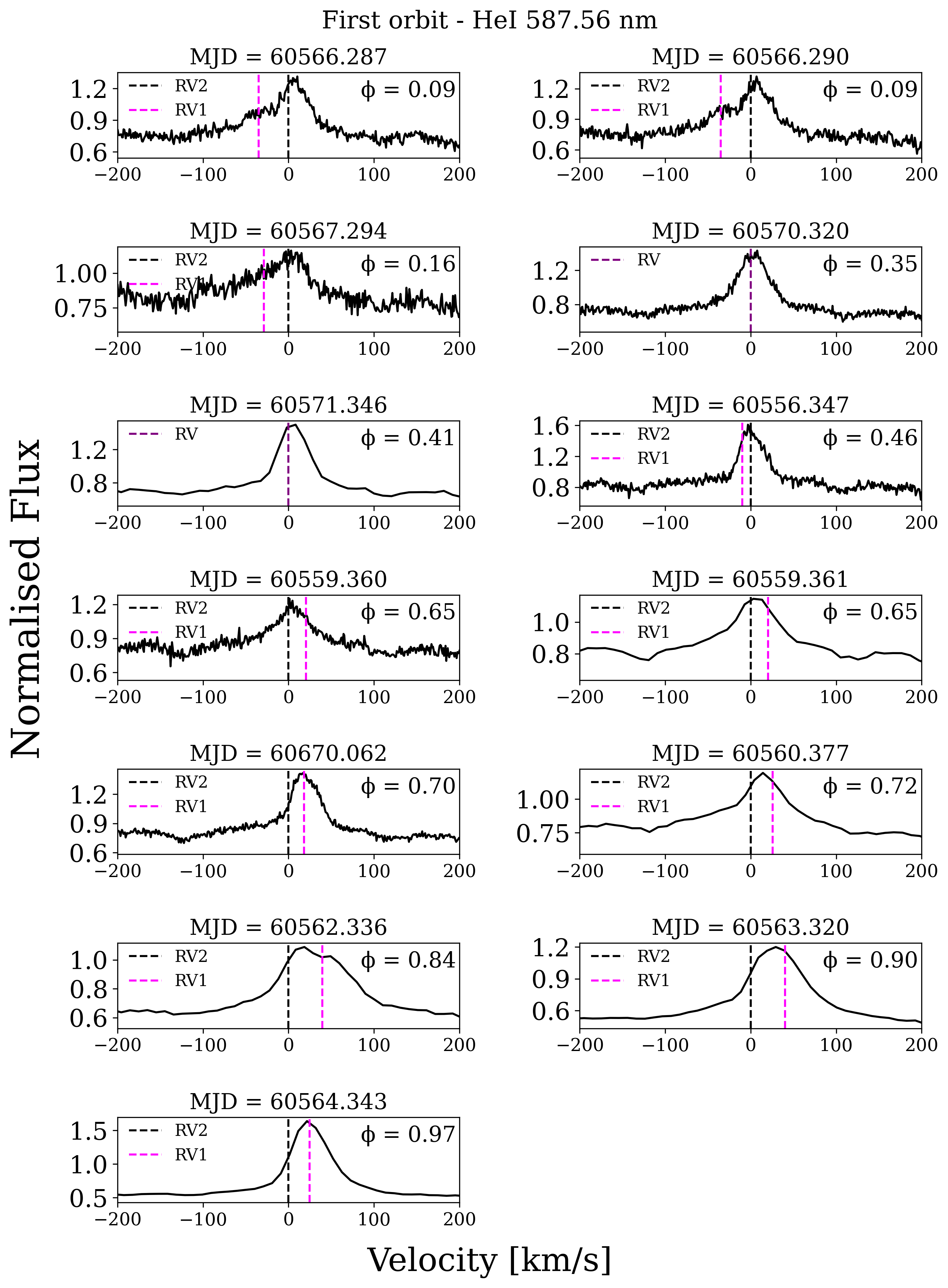 } 
    \caption{X-Shooter and UVES/HeI 587.56 nm line across the first orbit covered by this dataset. The orbital phase of each line is indicated in the right upper part of each sub-figure. The magenta and black dotted lines indicate the RVs of primary and secondary, respectively. Epochs with single RV value correspond to those with a single narrow-peaked BF ($RV_{1}=RV_{2}$) are indicated with purple dotted line. Lines are shifted to the rest-frame of the secondary.}
    \label{fig:HeI_first_cycle}
\end{figure}

\begin{figure*}
    \centering
    \includegraphics[width=\textwidth]{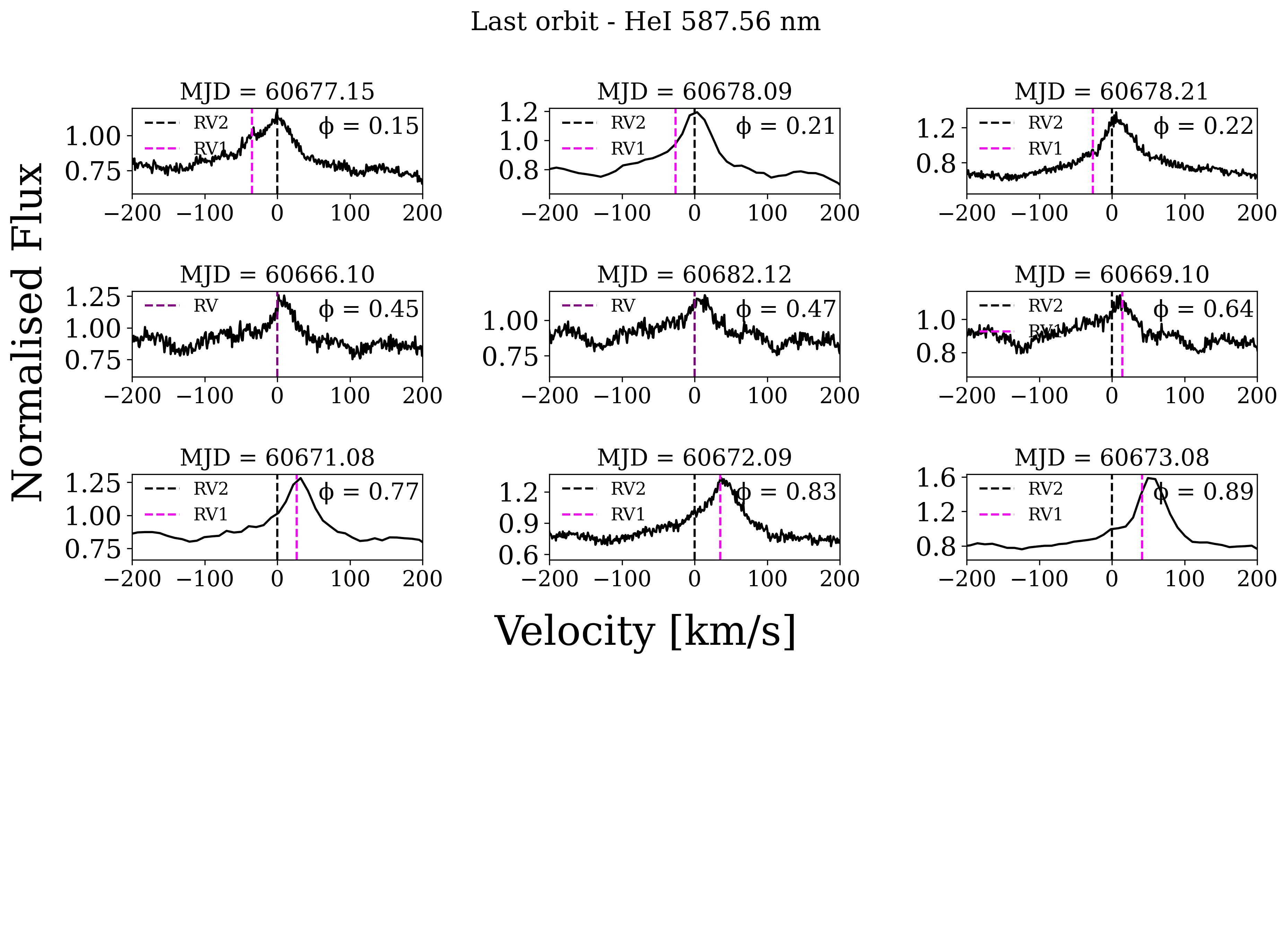 } 
    \vspace{-40mm}
    \caption{X-Shooter and UVES/HeI\, 587.56 nm line across the last two orbits covered by this dataset. The orbital phase of each line is indicated in the right upper part of each sub-figure. The magenta and black dotted lines indicate the RVs of the primary and secondary, respectively. Epochs with single RV value correspond to those with a single narrow-peaked BF ($RV_{1}=RV_{2})$ are indicated with purple dotted line. Lines are shifted to the rest-frame of the secondary.}
    \label{fig:HeI_last_2_cycles}
\end{figure*}

\begin{figure*}
    \centering
    \includegraphics[width=0.95\textwidth]{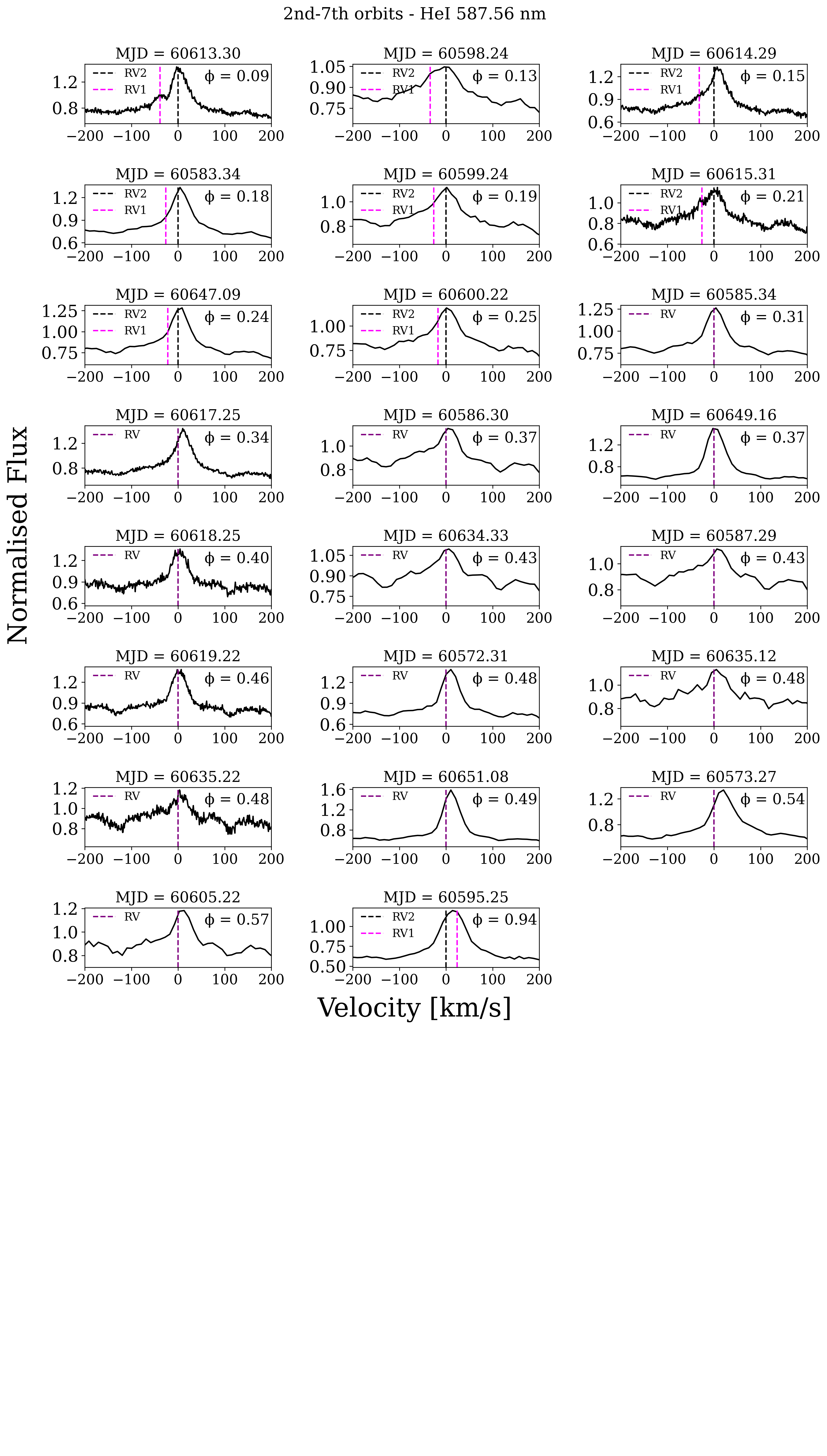} %

    \vspace{-88mm}
    \caption{X-Shooter and UVES/HeI\, 587.56 nm line across the middle orbital cycles. The orbital phase of each line is indicated in the right upper part of each sub-figure. The magenta and black dotted lines indicate the RVs of primary and secondary, respectively. Epochs with single RV value correspond to those with a single narrow-peaked BF ($RV_{1}=RV_{2}$) are indicated with purple dotted line. Lines are shifted to the rest-frame of the secondary.}
    \label{fig:HeI_middle/cycles}
\end{figure*}

 We show here the He I 587.56 nm line observed across the first, middle, and last orbital cycles of DQ Tau in the X-Shooter and UVES data (Fig.~\ref{fig:HeI_first_cycle}-\ref{fig:HeI_middle/cycles}).

\clearpage

\section{Log of observations}

\begin{table}[h!]
    
    \centering
\captionsetup{skip=10pt} 
    \caption{Log of the observations, radial velocity measurements and S/N of the reduced spectra for each spectrograph.}\label{table_all}
\resizebox{0.8\textwidth}{!}{\input{xshoo_uves_dqtau_mean_rv1_rv2_22_april_lateX_comp_sorted_by_mjd.tex}}
    \tablefoot{The RV values of both components are calculated using the BF technique on X-Shooter and UVES spectra. Dashed $RV_{1}$ values correspond to the epochs, where we can not distinguish the RV signature of both components. Therefore, we assume $RV_{1}=RV_{2}$ , i.e, the BF is single and narrow-peaked. In these cases, $RV_{1}$ values were assumed with higher errors compared to RV2 values.}

\end{table}

\clearpage
\section{Table of derived fluxes, equivalent widths, relative veiling factors, and accretion luminosities}

\begin{sidewaystable}
    \centering
    \small
    \caption{Derived integrated flux, equivalent width, relative veiling factor, and
             $\log{\left( \frac{L_{\mathrm{acc}}}{L_{\odot}} \right)}$ for each epoch.}
    \label{table_flux_ratios}

    \resizebox{0.9\linewidth}{!}{%
        \input{veiling_accretion_table_final.tex}
    }

    \tablefoot{Dashed values stand for the epochs where we have blended emission.
    Therefore, we derive single integrated flux and $log({{Lacc}}/{L_{\odot})}$ values
    added to the $F_{\mathrm{total}}$ and $log({Lacc_{\mathrm{total}}}/{L_{\odot})}$ columns.}
\end{sidewaystable}

\end{appendix}
\end{document}

%% file: xshoo_uves_dqtau_mean_rv1_rv2_22_april_lateX_comp_sorted_by_mjd.tex
\begin{tabular}{cccccccc}
\hline
Spectrograph & MJD & Obs date & $RV_{1}$ & $\sigma_{1}$ & $RV_{2}$ & $\sigma_{2}$ & S/N \\
\hline
UVES & 60556.3473 & 2024-09-03T08:20:02.582 & 19.472 & 1.339 & 29.448 & 1.017 & 13.08 \\
UVES & 60559.3603 & 2024-09-06T08:38:45.925 & 36.564 & 4.739 & 16.265 & 2.474 & 15.63 \\
X-Shooter & 60559.3611 & 2024-09-06T08:39:58.340 & 25.534 & 1.923 & 8.485 & 1.453 & 43.43 \\
X-Shooter & 60560.3768 & 2024-09-07T09:02:34.464 & 30.655 & 1.779 & 4.995 & 1.193 & 48.87 \\
X-Shooter & 60562.3363 & 2024-09-09T08:04:18.219 & 40.453 & 0.000 & 0.703 & 0.284 & 62.28 \\
X-Shooter & 60563.3197 & 2024-09-10T07:40:19.868 & 40.368 & 0.169 & 0.169 & 0.502 & 76.71 \\
X-Shooter & 60564.3425 & 2024-09-11T08:13:13.816 & 32.325 & 0.745 & 7.721 & 2.727 & 68.50 \\
UVES & 60566.2867 & 2024-09-13T06:52:46.690 & 3.703 & 0.425 & 38.787 & 0.176 & 30.52 \\
UVES & 60566.2901 & 2024-09-13T06:57:46.067 & 3.619 & 0.305 & 38.593 & 0.201 & 11.33 \\
UVES & 60567.2944 & 2024-09-14T07:03:58.888 & 6.159 & 0.878 & 35.036 & 0.662 & 10.27 \\
UVES & 60570.3200 & 2024-09-17T07:40:45.505 & - & - & 23.578 & 0.203 & 16.45 \\
X-Shooter & 60571.3463 & 2024-09-18T08:18:39.848 & - & - & 20.406 & 0.481 & 37.33 \\
X-Shooter & 60572.3054 & 2024-09-19T07:19:43.649 & - & - & 18.288 & 0.427 & 36.43 \\
X-Shooter & 60573.2743 & 2024-09-20T06:35:02.610 & - & - & 18.371 & 0.578 & 39.41 \\
X-Shooter & 60583.3431 & 2024-09-30T08:14:02.391 & 6.514 & 1.904 & 32.724 & 0.892 & 50.76 \\
X-Shooter & 60585.3402 & 2024-10-02T08:09:54.165 & - & - & 22.009 & 0.500 & 41.12 \\
X-Shooter & 60586.3031 & 2024-10-03T07:16:27.712 & - & - & 21.982 & 0.376 & 37.82 \\
X-Shooter & 60587.2900 & 2024-10-04T06:57:35.924 & - & - & 19.578 & 0.441 & 34.67 \\
X-Shooter & 60595.2489 & 2024-10-12T05:58:20.894 & 33.540 & 1.968 & 9.724 & 1.284 & 55.14 \\
X-Shooter & 60598.2419 & 2024-10-15T05:48:23.514 & 2.633 & 1.237 & 36.726 & 0.958 & 50.22 \\
X-Shooter & 60599.2394 & 2024-10-16T05:44:41.765 & 4.623 & 1.146 & 30.939 & 0.766 & 46.77 \\
X-Shooter & 60600.2194 & 2024-10-17T05:15:56.614 & 13.827 & 2.895 & 31.118 & 0.716 & 45.92 \\
X-Shooter & 60605.2246 & 2024-10-22T05:23:27.302 & - & - & 17.526 & 0.769 & 36.54 \\
UVES & 60613.3049 & 2024-10-30T07:18:59.902 & 2.828 & 0.685 & 41.352 & 0.265 & 23.86 \\
UVES & 60614.2878 & 2024-10-31T06:54:23.536 & 4.464 & 0.788 & 35.974 & 0.232 & 27.89 \\
UVES & 60615.3078 & 2024-11-01T07:23:14.125 & 6.717 & 0.963 & 32.555 & 0.294 & 26.92 \\
UVES & 60617.2498 & 2024-11-03T05:59:43.470 & - & - & 22.100 & 0.163 & 28.74 \\
UVES & 60618.2521 & 2024-11-04T06:03:04.510 & - & - & 22.771 & 0.313 & 17.34 \\
UVES & 60619.2213 & 2024-11-05T05:18:39.420 & - & - & 22.383 & 0.156 & 25.27 \\
X-Shooter & 60634.3325 & 2024-11-20T07:58:46.090 & - & - & 20.896 & 0.453 & 35.48 \\
X-Shooter & 60635.1223 & 2024-11-21T02:56:03.098 & - & - & 21.725 & 0.513 & 33.16 \\
UVES & 60635.2244 & 2024-11-21T05:23:09.887 & - & - & 22.021 & 0.258 & 22.18 \\
X-Shooter & 60647.0948 & 2024-12-03T02:16:30.919 & 10.631 & 1.728 & 32.713 & 0.830 & 47.57 \\
X-Shooter & 60649.1622 & 2024-12-05T03:53:31.891 & - & - & 21.950 & 0.587 & 44.46 \\
X-Shooter & 60651.0810 & 2024-12-07T01:56:37.127 & - & - & 18.661 & 0.605 & 40.59 \\
UVES & 60666.1012 & 2024-12-22T02:25:46.064 & - & - & 21.545 & 0.150 & 6.04 \\
UVES & 60669.1003 & 2024-12-25T02:24:25.592 & 27.483 & 0.710 & 13.235 & 0.685 & 27.63 \\
UVES & 60670.0615 & 2024-12-26T01:28:34.295 & 29.198 & 1.797 & 11.105 & 0.951 & 30.09 \\
X-Shooter & 60671.0782 & 2024-12-27T01:52:37.461 & 33.613 & 0.940 & 7.036 & 0.460 & 48.45 \\
UVES & 60672.0943 & 2024-12-28T02:15:45.179 & 39.173 & 1.088 & 3.726 & 0.611 & 17.24 \\
X-Shooter & 60673.0831 & 2024-12-29T01:59:35.525 & 40.367 & 0.172 & -1.125 & 0.425 & 57.79 \\
UVES & 60677.1474 & 2025-01-02T03:32:12.340 & 4.681 & 0.503 & 39.379 & 0.285 & 27.85 \\
X-Shooter & 60678.0910 & 2025-01-03T02:11:02.833 & 6.362 & 0.996 & 32.607 & 0.618 & 50.83 \\
UVES & 60678.2086 & 2025-01-03T05:00:23.899 & 6.139 & 0.568 & 32.354 & 0.193 & 18.88 \\
UVES & 60682.1153 & 2025-01-07T02:46:01.030 & - & - & 21.976 & 0.173 & 20.61 \\
\hline
\end{tabular}

%% file: veiling_accretion_table_final.tex
\begin{tabular}{ccccccccccccccc}

\hline

MJD & $F_{1} [\frac{\mathrm{erg}}{\mathrm{cm}^2 \cdot \mathrm{s}}]$
 & $F_{2} [\frac{\mathrm{erg}}{\mathrm{cm}^2 \cdot \mathrm{s}}]
$ & $F_{\text{total}}$ & $EW_{1}[nm]$ & $EW_{2}[nm]$ & $\sigma_{\mathrm{EW_1}}$[10$^{-3}$ nm]
 & $\sigma_{\mathrm{EW_2}}$[10$^{-3}$ nm]& $VF_{1}$ & $VF_{2}$ & $\sigma_{\mathrm{VF_1}}$ & $\sigma_{\mathrm{VF_2}}$ & $\log{\left( \frac{L_{\mathrm{acc,1}}}{L_{\odot}} \right)}$ & $\log{\left( \frac{L_{\mathrm{acc,2}}}{L_{\odot}} \right)}$ & $\log{\left( \frac{L_{\mathrm{acc,total}}}{L_{\odot}} \right)}$ \\
\hline
60559.361 & - & - & 1.580e-14 & 0.03 & 0.03 & 3.97 & 3.03 & -0.13 & 0.07 & 0.13 & 0.11 & - & - & -1.15 \\ 
60560.377 & 1.810e-15 & 1.350e-14 & 1.531e-14 & 0.02 & 0.03 & 0.18 & 0.16 & 0.01 & 0.03 & 0.01 & 0.01 & -2.20 & -1.23 & -1.18 \\ 
60562.336 & 7.420e-14 & 1.460e-14 & 8.88e-14 & 0.02 & 0.03 & 0.07 & 0.05 & 0.45 & 0.13 & 0.01 & 0.00 & -0.41 & -1.19 & -0.34 \\ 
60563.320 & 8.170e-14 & 8.920e-14 & 1.709e-13 & 0.01 & 0.02 & 0.18 & 0.17 & 0.86 & 0.51 & 0.03 & 0.01 & -0.36 & -0.32 & -0.04 \\ 
60564.343 & - & - & 1.160e-13 & 0.02 & 0.01 & 1.01 & 0.93 & 0.13 & 1.27 & 0.06 & 0.15 & - & - & -0.19 \\ 
60571.346 & - & - & 2.130e-14 & 0.02 & 0.03 & 0.03 & 0.03 & 0.00 & 0.08 & 0.00 & 0.00 & - & - & -1.01 \\ 
60572.305 & - & - & 2.520e-14 & 0.02 & 0.03 & 0.00 & 0.00 & 0.00 & 0.02 & 0.00 & 0.00 & - & - & -0.93 \\ 
60573.274 & - & - & 3.000e-14 & 0.02 & 0.03 & 0.04 & 0.04 & 0.00 & 0.09 & 0.00 & 0.00 & - & - & -0.84 \\ 
60583.343 & 6.020e-15 & 1.660e-14 & 2.262e-14 & 0.02 & 0.03 & 0.19 & 0.19 & 0.05 & -0.02 & 0.01 & 0.01 & -1.62 & -1.13 & -1.01 \\ 
60585.340 & - & - & 2.260e-14 & 0.02 & 0.03 & 0.01 & 0.01 & 0.00 & 0.12 & 0.00 & 0.00 & - & - & -0.98 \\ 
60586.303 & - & - & 1.930e-14 & 0.02 & 0.03 & 0.00 & 0.00 & 0.00 & -0.01 & 0.00 & 0.00 & - & - & -1.06 \\ 
60587.290 & - & - & 1.500e-14 & 0.02 & 0.03 & 0.00 & 0.00 & 0.00 & -0.05 & 0.00 & 0.00 & - & - & -1.18 \\ 
60595.249 & 5.440e-15 & 2.150e-14 & 2.694e-14 & 0.02 & 0.03 & 1.20 & 1.05 & 0.49 & 0.05 & 0.11 & 0.04 & -1.67 & -1.00 & -0.92 \\ 
60598.242 & 7.760e-15 & 1.290e-14 & 2.066e-14 & 0.02 & 0.03 & 1.03 & 1.22 & -0.00 & -0.01 & 0.04 & 0.04 & -1.49 & -1.25 & -1.05 \\ 
60599.239 & - & - & 9.700e-15 & 0.02 & 0.03 & 0.70 & 0.76 & 0.00 & -0.05 & 0.03 & 0.02 & - & - & -1.39 \\ 
60600.219 & 1.290e-15 & 1.650e-14 & 1.779e-14 & 0.03 & 0.03 & 1.45 & 1.57 & -0.14 & 0.13 & 0.05 & 0.06 & -2.36 & -1.13 & -1.11 \\ 
60605.225 & - & - & 1.440e-14 & 0.02 & 0.03 & 0.04 & 0.04 & 0.00 & -0.00 & 0.00 & 0.00 & - & - & -1.20 \\ 
60634.332 & - & - & 1.310e-14 & 0.02 & 0.03 & 0.05 & 0.05 & 0.00 & -0.03 & 0.00 & 0.00 & - & - & -1.24 \\ 
60635.122 & - & - & 1.730e-14 & 0.02 & 0.03 & 0.02 & 0.02 & 0.00 & -0.05 & 0.00 & 0.00 & - & - & -1.11 \\ 
60647.095 & 7.840e-15 & 1.140e-14 & 1.924e-14 & 0.03 & 0.03 & 1.21 & 1.38 & -0.09 & 0.12 & 0.04 & 0.06 & -1.49 & -1.31 & -1.09 \\ 
60649.162 & - & - & 2.740e-14 & 0.02 & 0.02 & 0.03 & 0.03 & 0.00 & 0.28 & 0.00 & 0.00 & - & - & -0.89 \\ 
60651.081 & - & - & 2.600e-14 & 0.02 & 0.03 & 0.00 & 0.00 & 0.00 & 0.25 & 0.00 & 0.00 & - & - & -0.91 \\ 
60671.078 & - & - & 2.050e-14 & 0.02 & 0.03 & 0.63 & 0.60 & 0.10 & -0.07 & 0.03 & 0.02 & - & - & -1.03 \\ 
60673.083 & 1.370e-14 & 1.800e-14 & 3.17e-14 & 0.02 & 0.03 & 0.13 & 0.11 & 0.34 & -0.02 & 0.01 & 0.00 & -1.22 & -1.09 & -0.85 \\ 
60678.091 & 8.500e-15 & 2.070e-14 & 2.92e-14 & 0.02 & 0.03 & 0.13 & 0.13 & 0.06 & -0.05 & 0.01 & 0.00 & -1.45 & -1.02 & -0.88 \\ 
\hline
\end{tabular}